\RequirePackage{rotating}
\documentclass[acmsmall]{acmart}

\usepackage{framed}
\usepackage{wrapfig}
\usepackage{siunitx}
\usepackage{enumitem}
\usepackage{threeparttable}
\usepackage{subfigure}
\usepackage{dcolumn}
\usepackage{amsmath}
\usepackage{rotating}
\usepackage{color}
\usepackage{xcolor}
\usepackage{multirow}

\newlength\MAX  
\newcommand*\Chart[1]{#1~\rlap{\textcolor{black!20}{\rule{\MAX}{2ex}}}\rule{#1\MAX}{2ex}}

\usepackage{tikz}
\usepackage{tikz-qtree}
\usetikzlibrary{trees}

\newcolumntype{d}{D{.}{.}{-1}}

\def\BibTeX{{\rm B\kern-.05em{\sc i\kern-.025em b}\kern-.08emT\kern-.1667em\lower.7ex\hbox{E}\kern-.125emX}}

\newcommand{\zhendong}[1]
{
   {\noindent\color{red}\bf [#1]$_{\scriptscriptstyle\textit{Zhendong}}$}
}
\newcommand{\yi}[1]
{
   {\noindent\color{blue}\bf [#1]$_{\scriptscriptstyle\textit{Yi}}$}
}
\newcommand{\yang}[1]
{
   {\noindent\color{green}\bf [#1]$_{\scriptscriptstyle\textit{Yang}}$}
}
\newcommand{\jim}[1]
{
   {\noindent\color{orange}\bf [#1]$_{\scriptscriptstyle\textit{Jim}}$}
}

\newcommand{\includeAuthorComments}[1]
{
   \ifthenelse{\equal{#1}{0}}
   {
      \renewcommand{\zhendong}[1]
      {
         {} 
      }
      \renewcommand{\yi}[1]
      {
         {} 
      }
      \renewcommand{\yang}[1]
      {
         {} 
      }
      \renewcommand{\jim}[1]
      {
         {} 
      }
   }{}
}

\AtBeginDocument{%
  \providecommand\BibTeX{{%
    \normalfont B\kern-0.5em{\scshape i\kern-0.25em b}\kern-0.8em\TeX}}}

\setcopyright{acmcopyright}
\copyrightyear{2019}
\acmYear{2019}
\acmDOI{10.1145/1122445.1122456}

\acmJournal{TOSEM}
\acmVolume{28}
\acmNumber{1}
\acmArticle{1}
\acmMonth{1}

\begin{document}

\title{Unveiling Elite Developers' Activities in Open Source Projects}

\author{Zhendong Wang}
\authornote{Both authors contributed equally to this research.}
\orcid{1234-5678-9012}
\affiliation{%
  \institution{University of California, Irvine}
  \streetaddress{Donald Bren Hall}
  \city{Irvine}
    \state{California}
  \postcode{92697-3425}
}
\email{zhendow@uci.edu}

\author{Yang Feng}
\authornotemark[1]
\affiliation{%
  \institution{University of California, Irvine}
  \streetaddress{Donald Bren Hall}
  \city{Irvine}
    \state{California}
  \postcode{92697-3425}
}
\email{yang.feng@uci.edu}

\author{Yi Wang}
\affiliation{%
  \institution{Rochester Institute of Technology}
  \streetaddress{70, 134 Lomb Memorial Dr.}
  \city{Rochester}
  \state{New York}
  \postcode{14623-5608}}
\email{yi.wang@rit.edu}

\author{James A. Jones}
\affiliation{%
  \institution{University of California, Irvine}
  \streetaddress{Donald Bren Hall}
  \city{Irvine}
    \state{California}
  \postcode{92697-3425}
}
\email{jajones@uci.edu}

\author{David Redmiles}
\affiliation{%
  \institution{University of California, Irvine}
  \streetaddress{Donald Bren Hall}
  \city{Irvine}
    \state{California}
  \postcode{92697-3425}
}
\email{redmiles@ics.uci.edu}

\renewcommand{\shortauthors}{Wang, Feng, et al.}

\begin{abstract}
Open source developers, particularly the elite developers, who own the administrative privileges for a project, maintain a diverse portfolio of contributing activities. 
They do not only commit source code but also spend a significant amount of efforts on other communicative, organizational, and supportive activities. 
However, almost all prior research focuses on a limited number of specific activities and fails to analyze  elite developers' activities in a comprehensive way.
To bridge this gap, we conduct an empirical study with fine-grained event data from 20 large open source projects hosted on \textsc{GitHub}. 
We investigate elite developers' contributing activities and their impacts on project outcomes. 
Our analyses reveal three key findings: 
(1) elite developers participate in a variety of activities while technical contributions (e.g., coding) account for a small proportion only; 
(2) elite developers tend to put more efforts into supportive and communicative activities and less efforts into coding as the project grows; 
and (3) elite developers' efforts in non-technical activities are negatively correlated with the project's outcomes in terms of productivity and quality in general, except for positive correlation with the bug fix rate (a quality indicator).
These results provide an integrated view of elite developers' activities and can inform an individual's decision making about effort allocation, thus might lead to finer project outcomes.
The results also provide implications for supporting these elite developers.
\end{abstract}



\maketitle

\section{Introduction}
\label{sec:intro}
Open source software (OSS) has become an engine for innovation and critical infrastructure for software development~\cite{Crowston:2008:FOS:2089125.2089127}.
OSS development is supported by communities formed from a loose collection of individuals. 
The contribution from these individual developers consists of various software-engineering activities, such as coding, bug fixing, bug reporting, testing, and documentation. 
All of these activities lead to the development and improvement of OSS projects, and fundamentally influence their outcomes.

Meanwhile, previous research, e.g.~\cite{Mockus:2002:TCS:567793.567795, crowston2005social, ducheneaut2005socialization, Jensen:2007:RMA:1248820.1248869, Crowston:2008:FOS:2089125.2089127}, has reported that among hundreds of such individuals, only a small portion of elite developers\footnote{
We use the term ``elite developers'' instead of the more commonly used ``core developers'' to refer to those who hold clearly defined project management privileges in a project, as opposed to \emph{only} being core \emph{code} contributors.
} contribute most of the code and oversee the progress of the project \cite{crowston2005social,Jensen:2007:RMA:1248820.1248869, LaToza:2006:MMM:1134285.1134355}. 
For example, in  Mockus et al.'s study on the Apache community~\cite{Mockus:2002:TCS:567793.567795}, they observed that the top 15 contributors (out of 388 total) had contributed over 83\% of modification requests and 66\% of problem reports. 
Furthermore, elite developers are also involved in many software-engineering activities beyond committing source code, such as moderating the discussions of an unfixed issue, documenting changes, organizing the project, and communicating with other contributors \cite{ducheneaut2005socialization}.
Therefore, analyzing the elite developers' activity is critical to understand the development of OSS projects.

Software developers maintain diverse activity profiles, including implementing new features, documenting changes and design, analyzing requirements, and fixing bugs~\cite{LaToza:2006:MMM:1134285.1134355}. Contributing source code is only one of such activities that an elite developer pursues. Prior studies each typically cast insights on one such specific non-coding activity, e.g., peer review \cite{Rigby:2008:OSS:1368088.1368162} or committing code \cite{daCosta:2014:UDC:2554850.2555030}. Most fall short of providing an integrated view on all of the developers' activities and the distribution of efforts on these activities. Even though these studies provide guidance to software developers on improving some software engineering tasks, such as assigning bug reports \cite{shihab2013studying, guo2011not} and estimating cost \cite{Amor:2006:EEC:1139113.1139116}, we cannot fully realize the activity data to inform better decision-making and ultimately bring better project output without a comprehensive study of a diverse range of developer activities including their communicative, organizational and supportive activities which are beyond typical technical ones. Because of these activities, together with typical technical activities, influence the software systems being developed in different ways, understanding the elite developers' activities, beyond coding, draws the most critical development knowledge and experience from the community. 
This leads to our first research question:

\begin{description}[leftmargin=1em]
\item [$\mathbf{RQ_1}$:] \emph{What do elite developers do in addition to contributing typical technical code in OSS projects?}
\end{description}

Since software engineering is a human-centered activity~\cite{flegal2008overthinking}, effectively managing human resources may significantly enhance project productivity and collaboration quality. However, it is not clear how elite developers change their activities and which kind of tasks they focus on the development of OSS projects. Understanding the dynamic evolution of elite developers' effort distributions over different activity categories over the life cycle of OSS projects can guide the behavior of junior developers and also assist resource management.

This gives rise to our second research question:
\begin{description}[leftmargin=1em]
\item [$\mathbf{RQ_2}$] \emph{How do an OSS project's elite developers' effort distributions evolve along with the growth of the project?}
\end{description}

Given that OSS projects are developed by elite developers as well as many external contributors, elite developers' activities, especially the ones beyond technical contributions, such as communicating with bug reporters, the documenting project changes, assigning tasks and labeling issues, may fundamentally influence the outcome of the whole team. Because successful software engineering activities require qualified developers with the proper expertise to complete the task efficiently and effectively, understanding these impacts are critical for developers to oversee the project for assuring the development productivity and product quality.
Thus, we have our third research question:
\begin{description}[leftmargin=1em]
\item [$\mathbf{RQ_3}$] \emph{What are the relationship between an OSS project's elite developers' effort distributions and the project's outcomes in terms of productivity and quality?}
\end{description}

To answer the above research questions, we conduct an empirical study using fine-grained event data from 20 large open-source projects hosted on \textsc{GitHub} consisting of both company-sponsored and non-company-sponsored projects. To better utilize the activity data to draw insights about the elite, we first map them from raw atomic events to sense-making high-level categories. These categories are: \emph{\textbf{communicative}}, \emph{\textbf{organizational}}, \emph{\textbf{supportive}}, and \emph{\textbf{typical}}. Their detailed definitions and mapping protocols are introduced in Section 3.3. We then use multiple techniques to model and analyze the data. Our study reveals three main findings. First, elite developers participate in a variety of activities, while coding only accounts for a small proportion. 
Second, with the progress of the project, elite developers tend to be increasingly involved in more non-technical activities, while decreasing their coding and other technical activities. 
Third, elite developers' effort distributions exhibit complex relationships with project productivity and quality. For both project productivity indicators (no. of new commits and average bug cycle time in each project-month), our results suggest that project productivity has negative correlations with efforts in non-technical (communicative, organizational, and supportive) activities. For one project quality indicator (no. of new bugs in each project-month), our results show that project quality has negative correlations with  efforts in non-technical activities; however, for the other project quality indicator (bug fix rate in each project-month), our results show that project quality has positive correlations with efforts in supportive activities.

The main contributions of this article are three-fold. 
\begin{itemize}
    \item We conduct an empirical study that not only characterizes elite developers' activities and their dynamics, but also identifies the relationships between elite developers' activities and project outcomes. 
    Based on the findings, we identify a set of actionable recommendations for practitioners.
    \item We take a fresh perspective to investigating the activities of OSS developers through collecting, modeling and analyzing all kinds of publicly available online software-engineering activities of developers rather than focusing on one or several specific activities, and thus obtain a holistic view of the OSS development.
    \item We set up a \textit{well-cleaned} dataset comprising all the event data of large OSS projects, which is made publicly available\footnote{https://drive.google.com/drive/folders/10ibmz2svPRf3jfRtm7mbiouo9ATaYAoB}. 
\end{itemize}

Our work is built on SE communities' continuous efforts in investigating and assisting OSS projects in the last two decades. Researchers have investigated community structures and compositions, individual motivation, behavior and experiences, as well as these factors' impacts, e.g., \cite{bird2011don, jergensen2011onion,qiu2018going, rahman2011ownership, valiev2018ecosystem, vasilescu2015gender, vasilescu2016sky}. While these extant studies build solid knowledge on OSS projects, most of them focus on the code-contribution-related activity, such as coding, reviewing, testing, debugging, and so on. Our work expands the literature by enhanced understandings about the breadth and dynamics of elite developers' activities, and their correlation with project outcomes.

\section{Background \& Related Work}
In this section, we briefly overview the backgrounds of this study and discuss the related work about open source communities and developer's activities. We start from a brief introduction of the hierarchical structure of open source communities, followed by discussions of relationships between developers and the activities of developers. We also highlight how current work distinguishes itself from the prior.

\subsection{The Hierarchical Open Source Community}

Open source development has been a mainstream practice in building modern software systems \cite{Crowston:2008:FOS:2089125.2089127}. Different from traditional development paradigms, an open source project is centralized on its community that produces collective goods through collaboration among its members \cite{howison2014collaboration}. Though the detailed governance and social practices may vary in different projects \cite{o'mahony2007the}, members of an open-source community usually have different roles regarding their responsibilities, rights, and levels of contributions \cite{Scacchi:2007:FSS:1287624.1287689, Bird:2008:LSS:1453101.1453107}. 
Similar to other hierarchical organizations, an OSS community follows an onion-shaped social structure \cite{crowston2005social}.

There are several different definitions for each layer in this hierarchical community \cite{Jensen:2007:RMA:1248820.1248869, crowston2005social, ducheneaut2005socialization}, but in general there are five major types from core developers, internal and external contributors, issue reporters, and finally to peripheral users (note that terms may differ from study to study). However, members in a project may have several statuses with more detailed differences.

Peripheral users of an OSS project usually are users of the software artifacts, but never contribute to the project directly (other than sending user feedback or usage data). For most users of an OSS project, a peripheral user is the starting point unless they have achieved recognition in the same ecosystem \cite{jergensen2011onion}.
If these peripheral users wished to contribute to more critical tasks of the project, they usually have to get through a socialization process. In Ducheneaut's case study \cite{ducheneaut2005socialization}, he reveals the socialization path of becoming a core developer when starting at the periphery. This path includes socialization with the current core team, and completing a series of development tasks from simple to complicated. After being socially recognized by experts for a project, they join the core team and become core developers, themselves. Thus, they are granted privileges of this project (i.e., get project ``tenure'' in a repository). Further, they start to have the administrative power in the project; for example, they can oversee other external contributors' technical submissions.

Current OSS development, especially large-scale projects, can be described under the ``umbrella'' of an ecosystem. In a follow-up study, Jergensen et al. discussed the evolution of this socialization process in the context of modern open-source-software development \cite{jergensen2011onion}. As the technologies developed for software engineers, such as advances in version-control systems (\textit{git}), fewer open-source projects are being developed solely in isolation. Further, more projects are developed in parallel under the broader context of software ecosystems. In their study, they found that there are several types of contributors among open-source users across different projects \cite{iannacci2005coordination}. In addition, many developers move from project to project like ``nomads.'' Another critical finding is that, in an OSS project, as developers gain more technical experience, their contribution is not towards the core of the project in terms of code centrality.

Among many studies on OSS communities, researchers have come to the consensus that only a small portion of developers make the most contribution \cite{crowston2005social,Jensen:2007:RMA:1248820.1248869, LaToza:2006:MMM:1134285.1134355}. Understanding elite developers is critical in investigating the health and sustainability of the community, and various methods have been employed to analyze their activities. Meanwhile, please note that members have developed a shared basis of authorities and privileges in most of mature open source projects (``bazaar'' in Eric Raymond's ideology \cite{raymond1999the}), and enabled transferring authorities and privileges among them. Thus, a member's identity of ``being an elite'' is indeed dynamic \cite{o'mahony2007the}.

\subsection{The Role-Based Relationships among Developers}

Members of a software project form complex relationships in a wide range of development activities. Tab. \ref{tab:role} shows a brief overview of studies investigating the relationships between developers. While ``core vs. peripheral'' is a dominant terminology used in SE literature to characterize role-based developers' relationships, there are also other terminologies being proposed. For instance, when outside or novice developers are in the process of joining and learning from an existing project, the \textit{mentor} and \textit{newcomer} relationship between developers would eventually be established for social and technical reasons \cite{Dagenais:2010:MNS:1806799.1806842, Canfora:2012:GMN:2393596.2393647}. Prior research also suggests that the mentorship activities face several barriers for both ends of stakeholders \cite{Balali2018}, which is threatening the sustainability of the project. In another scenario, when software companies open source their internal software to the public domain, understanding the contribution behaviors and the impacts of \textit{external} and \textit{internal} contributors for these company-sponsored projects become critical for many purposes. For example, companies might want to find a balance of management efforts and fast iteration of enhancement when receiving external help. As we mentioned before, developers' roles have a temporal characteristic. Role migrations are very common \cite{jergensen2011onion}. A peripheral member could be promoted to a core member; a newcomer could be a mentor after his/her skills have developed. However, such a growing process may take substantial effort. For instance, external members gain the roles of internal members could be very painstaking \cite{dias2018drives, ducheneaut2005socialization}. Besides, role migrations are not single directional. For example, core members may lose their roles if they no longer actively contribute to the project \cite{lin2017developer}. Therefore, role-based relationships are dynamic in nature.

\begin{table}[]
\small
\caption{Comparisons between elite developers and other similar roles defined in literature.}
\begin{tabular}{lp{2.8cm}p{2.2cm}p{6cm}}
     
\toprule
\textbf{Role}               & \textbf{Complement Role} & \textbf{Context} & \textbf{Role Definition} \\
\midrule
Core     & Peripheral & Open-source & Members of a small core group who are mainly responsible for overseeing and contributing to the project \cite{Mockus:2002:TCS:567793.567795, crowston2005social, Crowston:2008:FOS:2089125.2089127}.           \\
\midrule
Maintainer         & Contributor & Open-source & Members who are responsible for a software module, mainly in accepting contributed patches \cite{Jensen:2007:RMA:1248820.1248869, ducheneaut2005socialization}.           \\
\midrule
Internal & External & Company-sponsored Open-source & Individuals who are members of the development group; usually are marked as contributors on the project homepage \cite{wagstrom2012roles, wagstrom2009vertical, dias2018drives}.        \\
\midrule
Mentor             & Newcomer & General software development & Persons who train and help novice and inexperienced (newcomer) developers for project details \cite{Balali2018, Dagenais:2010:MNS:1806799.1806842, Canfora:2012:GMN:2393596.2393647}.           \\
\midrule
Elite              & Non-elite & Open-source & Developers who own administrative privileges in the project [this study].   \\
\bottomrule          
\end{tabular}
\label{tab:role}
\end{table}

Open-source software and its developer community have been substantially evolving since the 2000s. The transparency afforded by online open-source code hosting sites such as GitHub enables researchers and practitioners to closely observe and review the dynamics of the projects, such as the evolution of software artifacts, the trajectories of peripheral participants' self-development \cite{dabbish2012social}. Meanwhile, supported by versioning control systems and logs from various communication channels, a tremendous volume of social and technical latent data can be acquired \cite{bird2011sociotechnical}. Various empirical research has been postulated to obtain valuable knowledge from these datasets for software design, development, and quality assurance.

\subsection{The Activities of Developers}
In an early study, organizational psychologist Sonnentag conducted an empirical study with software-company professionals to study their weekly activities in software development \cite{sonnentag1995excellent}. Based on her observations and the grounded theory process, she classified four broad types of activities in the professional lives of developers. The broad types are \textit{communicative}, \textit{organizational}, \textit{supportive}, and \textit{typical}. Further, based on her field study with excellent and average software professionals, she found excellent and average developers usually spend a similar amount of effort on typical software-engineering tasks such as coding, testing, and debugging. However, excellent developers spend more time on meeting and consulting. This study is critical in identifying the comprehensive set of software engineers' activities. However, this study is limited with a specific company context, and may be not suitable to describe open source development. In addition, they have not investigated the impact of additional activities and the burdens of elite developers.

Later in another open-source study, Wagstrom et al. classified the roles of open-source contributors. Besides five typical types of users, they also classified special roles in the ecosystem development based on their code-related contribution \cite{wagstrom2012roles}, such as, ``code warrior'' who continuously contributes to the project by submitting commits, and ``project rockstar'' who also submits a tremendous amount of code and also has very high community exposure in terms of follower numbers. 
This study categorizes OSS developers based on their code contribution activities. They employ the milestone-event for categorizing users into five major hierarchical layers, and define special roles for ecosystem-scale development. In their role classification, they consider code-related activities such as submitting source code or reporting bugs.

In a recent study of software-development expertise, Baltes and Diehl~\cite{baltes2018towards} conducted surveys on 335 software developers who are active over \textsc{GitHub} and \textsc{StackOverflow}. Based on the survey result, they created a theory to describe important factors influencing the experts' performance. Their work is critical in providing a theoretical lens for software engineer's expertise, but it is limited to experience of programming (typical software-engineering activities). Further, their results rely on a self-reported survey without empirical verification.

\section{Empirical Study Design}
To answer the three research questions presented in Section~\ref{sec:intro}, we conduct an empirical study based on 20 open-source projects. This section introduces the design of the study.

\begin{table}[!h]
\small
    \centering
    \begin{threeparttable}
     \caption{Sample projects and their Description.\label{tab:sample_projects}}
    \begin{tabular}{ll}
    \toprule
    \textbf{Project}& \textbf{Description}\\
    \midrule
    Aframe        & Web framework for virtual reality applications \\
    Alamofire     & Swift library for HTTP networking \\
    ExoPlayer*    & Media player for Android \\
    Finagle*      & Extensible RPC system for JVM \\
    Fresco*       & Android library for images \\
    Guava*        & Set of various Java libraries \\
    Immutable-js* & JavaScript library for immutable data structure\\ 
    Jest*         & JavaScript testing framework \\
    Marko*        & JavaScript library for building UI\\
    Moya          & Swift network framework \\
    Nightmare*    & Browser automation library\\
    Rclone        & Program to sync files\\
    React*        & JavaScript library for building UI \\
    Recharts      & JavaScript chart library \\
    Sqlitebrowser & Visual UI for databases in SQLite\\
    Stf           & Smart device testing framework \\
    Tensorflow*   & Library for numerical computation \\ 
    Tesseract     & Text recognition (OCR) engine \\
    Tidb*         & Distributed database system \\
    ZeroNet       & Decentralizes websites to be resistant to censorship\\
    \bottomrule
    \end{tabular}
    \begin{tablenotes}
    \item *: Projects sponsored by companies.
    \end{tablenotes}
    \end{threeparttable}
\end{table}

\subsection{Targeted Projects}

We select 20 open-source projects hosting their repositories on \textsc{GitHub} as the targets of the study. Tab.~\ref{tab:sample_projects} lists them with short descriptions.
The selection of the targeted projects is not random. They are selected based on four considerations. First, the selected projects are all large projects that have established administration structures (having some forms of the formal project management committee and soliciting contributions through the pull-request model). They must be large enough and have traceable records of continuous contributions from a set of contributors (at least 100 pull-requests and 50 contributors historically). Second, the selected projects represent a diverse sample of projects in terms of application domains, such as a testing framework (jest), a popular deep-learning library (Tensorflow), a multi-media player (ExoPlayer), a web-development framework (React), and a database (Tidb). Third, our sample includes a subset of company-sponsored ($n=11$) projects, which reflects the trend of the increasing involvements of companies in open-source development \cite{wagstrom2009vertical}. Last, the sampled projects should maintain a relatively long traceable records on \textsc{GitHub}, which allows us to study the longitudinal dynamics while keeping the data consistency.

\subsection{Data Preparations}
\label{sec:datapreparation}
\begin{figure}[!h]
     \centering
     \includegraphics[width=\linewidth]{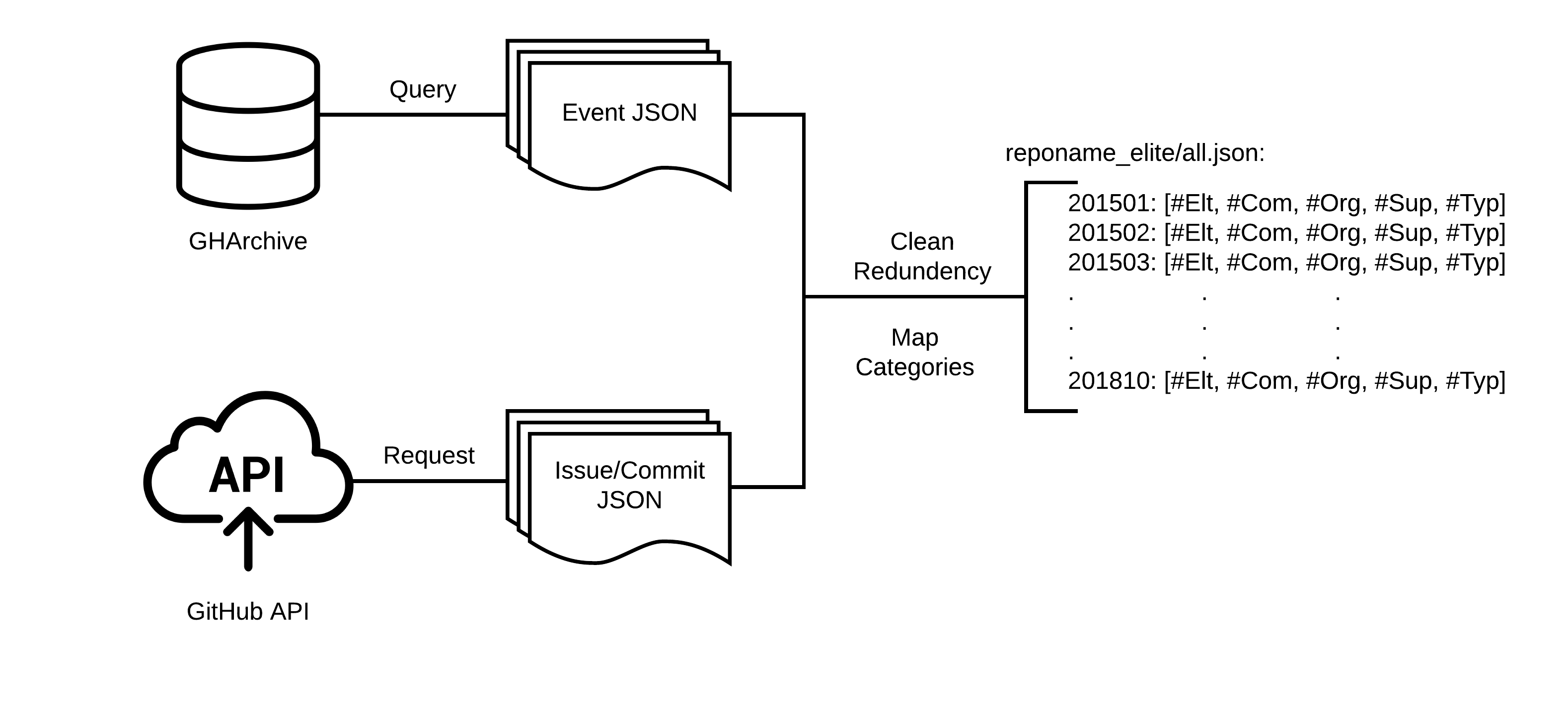}
     \vspace{-.5in}
     \caption{Data collection and cleanup process.}
     \label{data}
 \end{figure}

The current version of the \textsc{GitHub} API only allows us to retrieve 300 events or events from the past 90 days, whichever met first\footnote{\textsc{GitHub} Event API: \url{https://developer.github.com/v3/activity/events/}}.
Therefore, in order to extract event data from a extended range of projects' lifecycle, we employ the \textsc{GitHubArchive} public data dump on Google Cloud. We also employ Google BigQuery to extract the monthly event log for each sampled repository from January 2015 to October 2018. Additionally, for repositories that started or were made public during the year 2015, we store data files starting from the project creation month.   
Fig. \ref{data} provides an overview of the data collection and cleaning process.

For each month, GHArchive provides most event logs on a repository such as push, open issues, open pull request, Gollum (editing wiki), and comments. We use SQL-like queries (designed by BigQuery) to search for projects, and save the results into tables of the personal Google Cloud database. Further, we export tables as JSON files to the cloud storage and download them to a local computer for later analysis. 
In total, we collected 5.60 GB of 900,862 events (communicative: 238,986; organizational: 42,317; supportive: 514,957; and typical: 104,602)  for these 20 repositories.

However, there are several types of events associated with issues that could not be recorded using the above method, e.g., assigning a ``won't fix'' label to an issue by a project administrator, or delegating a developer to investigate a newly posted bug. To fix this problem, we resort to a customized Python script via the \textsc{request}\footnote{Simplified HTTP request client for Python: https://github.com/request/request} library to request events from the \textsc{GitHub} API, and then to download issue event logs for every issue that has been reported in each repository. Thus, we collect precise project management information, such as who has the administrative privilege on a repository and oversees the progress of the project. In order to search for commits by date and author more easily, and derive necessary metrics for the later data analysis on the project productivity, we also download the commit logs of all sampled projects. In total, we have collected 1.81 GB data of issue events and commit logs.

Finally, we use Python scripts to merge event data based on event ID and commit SHA, and clean the redundant data that were recorded on both data sources. By using two data sources, there are some categories of events that were kept recording on each data source, such as \textit{close issue} and \textit{reopen issue}. Because the GHArchive project employs \textsc{GitHub} event API to archive activities on a daily basis, we decided to keep events from \textsc{GitHub} Issue API. We convert event logs into a monthly list based on the number of events that have happened in each major category.

\begin{figure}
\begin{tikzpicture}[grow'=right,level distance=1.25in,sibling distance=.25in, scale = .6]
\tikzset{
    root/.style={rectangle,draw,fill=blue!20},
    edge from parent/.style= 
        {thick, draw, edge from parent fork right},
    every node/.style=
        {minimum width=1in, text width=1in, align=center},
    every parent/.style=
        {draw, minimum width=1in, text width=1in, align=center}
    }

\begin{scope}[yshift=3cm]
\Tree 
    [. \large{\textsc{\textsf{Communicative}}} 
        [.{Comment on commit or issue}
                [. \texttt{CommitComment Event} ]
                [. \texttt{ContentReference Event} ]
                [. \texttt{IssueComment Event} ]
        ]
        [. {Edit comment}
                [. \texttt{Edit Event} ]
        ]
    ]
\end{scope}

\begin{scope}[yshift=-3cm]
\Tree
    [ . {\large{\textsc{\textsf{Organizational}}}}
        [. {Assign someone to issue or PR}
            [. \texttt{IssueAssign Event} ]
        ]
        [. {Manage code reviewer}
            [. \texttt{ReviewDismissed Event} ]
            [. \texttt{ReviewRequested Event} ]
        ]
        [. {Member Management}
            [. \texttt{OrgBlock Event} ]
            [. \texttt{Organization Event} ]
            [. \texttt{Member Event} ]
            [. \texttt{Team Event} ]
        ]
    ]
\end{scope}

\begin{scope}[yshift=-11cm]
\Tree
    [. \large{\textsc{\textsf{Typical}}}
        [. {Commit source code}
            [. \texttt{Commit Event} ]
        ]
        [. {Pull Request}
            [. \texttt{PullRequest Event} ]
            [. \texttt{PullRequestReview Event} ]
            [. \texttt{PullReuqestReview Comment Event} ]
        ]
    ]
\end{scope}

\begin{scope}[xshift=10.5cm, yshift=-1cm]
\Tree
    [. \large{\textsc{\textsf{Supportive}}}
        [. Document
            [. \texttt{Gollum Event} ]
            [. \texttt{(Un)label Event} ]
            [. \texttt{Release Event} ]
        ]
        [. Maintenance
            [. Branch
                [. \texttt{Create Event} ]
                [. \texttt{Delete Event} ]
                [. \texttt{Push Event} ]
            ]
            [. {Issue and PR}
                [. \texttt{Issue Event} ]
                [. \texttt{Issue(Un)Lock Event} ]
                [. \texttt{Merge Event} ]
                [. \texttt{IssueRename Event} ]
                [. \texttt{IssueDuplicate Event} ]
            ]
            [. Milestone
                [. \texttt{(De)milestone Event} ]
                [. \texttt{Project Event} ]
                [. \texttt{ProjectColumn Event} ]
                [. \texttt{ProjectCard Event} ]
            ]
            [. Checks
                [. \texttt{CheckRun Event} ]
                [. \texttt{CheckSuite Event} ]
            ]
        ]
    ]
\end{scope}
\end{tikzpicture}
\caption{The taxonomy of \textsc{GitHub} event types. The definition of each raw event can be found in the official \textsc{GitHub} Events API documentation page: \url{https://developer.github.com/v3/activity/events/}.}
\label{taxonomy}
\end{figure}
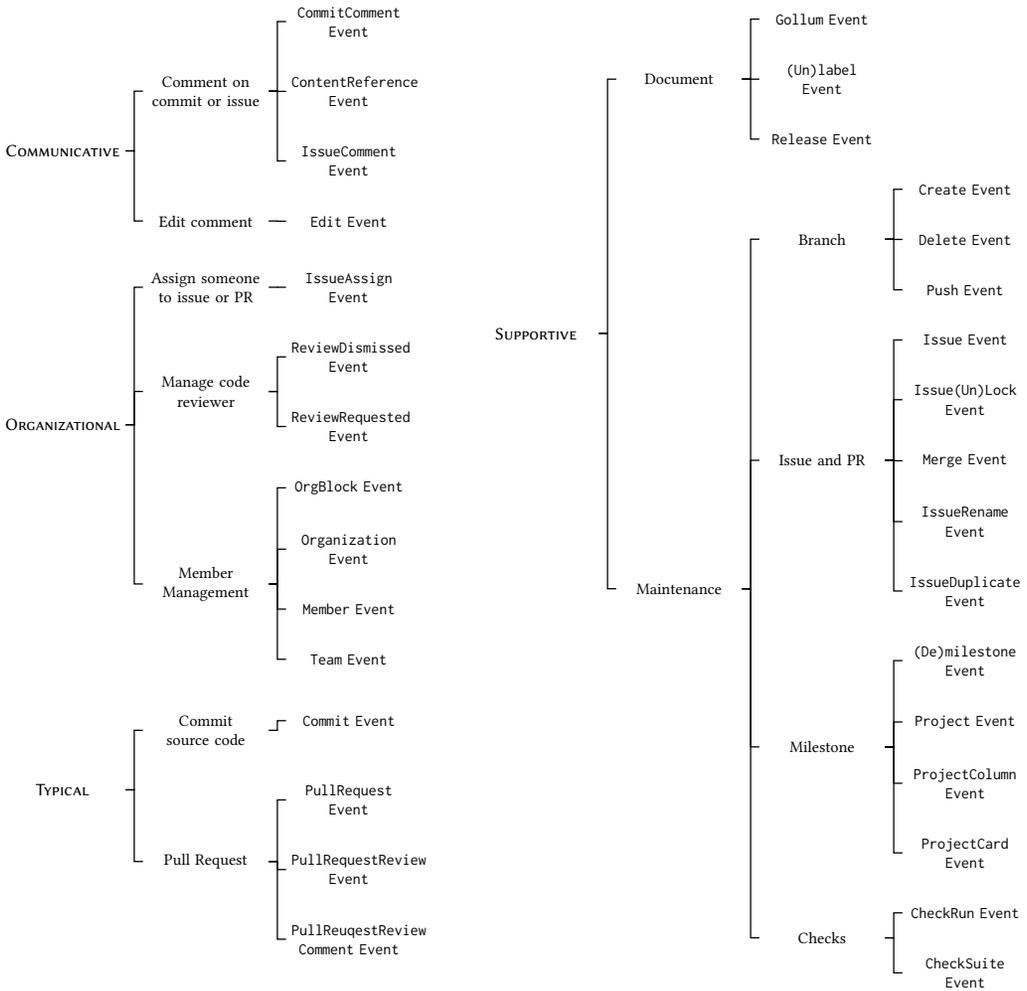


\subsection{Event Categories and Mapping}

Although the event log data faithfully records developers' activities, we need to recode the data unit categories that are easier for humans to understand and analyze. Particularly, the percentage of types for a developer group should be able to reflect the effort allocation and focused roles.

Collecting low-level activities from self-reported and observed data in the field, and then inductively mapping these activities onto broad categories for systematically extract behavior patterns and analyze work effort allocation, is common practice to establish the activity profile of a certain group \cite{brodbeck1994software, sonnentag1995excellent, sonnentag1998expertise, wynekoop2000investigating}. In one of prior field interview studies \cite{brodbeck1994software}, the work focuses and daily activities category of professional software developers were summarized based on subjects reported activities. Because we are particularly interested in investigating the overall activity profile of elite developers, we choose to follow an established category system that reflects the daily activity, instead of other low-level tasks based category system which focusing on coding activities.

We reuse the categorizing system created in Sonnentag \cite{ sonnentag1995excellent}, and to further investigate the contribution of elite developers, as well as the relationships between their work focuses and project outcomes for open-source projects. This study summarizes and categorizes professional software developers' daily activities into four major categories: \textit{communication}, \textit{organization}, \textit{support} and \textit{typical}. In their study, research subjects were developers in private companies; in order to make these definitions fit the context of open source development, we slightly modify the definition and operationalization of each category.

\subsubsection{Communicative} In the conventional co-located software development team, communicative activities usually refer to formal and informal meetings and consultations \cite{sonnentag1995excellent}. However, under the setting of distributed software development where open-source project usually employs, each project applies various communication channels including mailing list, instant message, and online discussion board \cite{bird2011sociotechnical}. Moreover, some projects such as Tensorflow, even apply other broadcast channels such as a blog, website, and YouTube channel. Therefore, similar to other empirical studies with open-source developers, we are not able to collect communicative activities on all channels. For example, private instant messages are often unavailable. However, as \textsc{GitHub} is the major platform for developers to exchange ideas, by extracting communicative event logs from \textsc{GitHub}, we are able to capture all \textit{public communicative traces} that happened on this platform by each contributor. 
    
The definition of communicative activities is \textit{public and visible communication through commenting features supported by the platform on issues, commit, and project milestones}.

\subsubsection{Organizational} In previous field studies, organizational activities are categorized as delegating tasks among the development team and other project organizations in professional software development. Thus, similarly under the open-source development settings, representative activities of this type are \textit{assigning} and \textit{unassigning} tasks to a developer, such as assigning someone with a GitHub issue or reviewing a pull request. 
    
We define organizational activities as \textit{managing the project community and delegating tasks, including code reviewing, debugging, and user support to internal and external contributors through the features supported by the platform}.

\subsubsection{Supportive} Supportive activities are critical to open-source development and mainly refer to other non-coding activities in collaboration. It includes \textit{documentation} work such as writing documentation/wikipage and categorize issues by adding labels to them. Further, supportive also includes \textit{maintenance} work, for example, managing development branches and release or archive code versions.

We define supportive activities as \textit{non-coding activities in the collaborative open-source development through techniques that are supported by the platform, including documentation, versioning control, and development branch management}.

\subsubsection{Typical} Typical activity in software development are coding, testing, debugging and reviewing on an \textit{individual} basis. Thus, under the setting open-source platform, we only include commit activity under this category. In addition, we count event actor as the commit \textit{author} rather the \textit{committer}, since the \textit{author} is the original developer who wrote the code.

We define typical activities as \textit{conventional code-writing task finished at the individual level, and counted as submitted commits and pull requests}.
 

\subsubsection{Mapping raw events to the above categories} 


We apply the closed card sorting method to place 35 raw \textsc{GitHub} events in these four major categories. Three researchers participate in the card sorting activity. Among three researchers, the card sorting yields 0.92 average joint probability of agreement. Especially, it achieves 82.8 \% relative observed agreement, and reaches 0.77 kappa \cite{brennan1981coefficient}. All differences among card sorters are discussed and resolved. The final mapping is shown in Fig.~\ref{taxonomy}.

By mapping the low-level raw \textsc{GitHub} events into these four categories, we can reason developers' activities at the level that makes sense to understand them as real work practices and human efforts in organizational settings \cite{pentland2005organizational} instead of losing in millions of tiny events which are not considered as an integrated work practices. Besides, we argue that such a categorization system precisely and comprehensively reflects the general work practices of professional software developers. The categorization system is borrowed from the literature \cite{brodbeck1994software, sonnentag1995excellent} of empirical field observations and interviews with a large number of software development projects and hundreds of professional developers. 
Although our study focuses on open-source developers, the types of their routine work practices at the individual-level in software development would be unlikely to go beyond the in-house software development, while the way of organizing such practices may be different at the collective-level \cite{Crowston:2004:EWP:1029997.1030003, gousios2016work, rolandsson2011open, von2006promise}. Doing so enables us to better study the dynamics of elite developers' work practices and their impacts, thus deriving meaningful findings and implications.

\subsection{Collecting Project Outcomes Data: Productivity and Quality}
Since one of the research goals is to investigate the impact of elite developers' activity on project outcomes ($\mathbf{RQ_3}$), we need to collect project-outcomes data. We consider two project outcomes: productivity and quality, which are viewed as the most important project outcomes \cite{vasilescu2015quality}. Each of them has two indicators, which are introduced as follows.

For productivity, the first indicator is: \emph{the number of a project's all new commits in a project-month\footnote{To simply the following discussion, we use the term ``project-month'' to denote \emph{a given month in a project}.}}. Thus, for project $i$ in month $m$, we use $NewC_{im}$ to denote it. In many studies focusing on the OSS development and community, the number of commits is considered as the productivity metric~(e.g., \cite{vasilescu2015quality, vasilescu2013stackoverflow, vasilescu2015gender}). Therefore, we adopt this widely-used productivity indicator. Note that we count the commits from all contributors rather than from elite developers only, because we measure the impact on the productivity of the whole team. The second indicator is \emph{the average cycle time of a project's closed bugs in a project-month}. Similarly, for project $i$ in month $m$, we use $BCT_{im}$ to denote it. Such an indicator has been used to measure project productivity in a bunch of prior studies (e.g., \cite{4228638,kim2006long}). 

Following the conventions in previous SE literature (e.g., \cite{khomh2012faster,vasilescu2015quality, ray2014large}), we first operationalize the code quality by \emph{the number of bugs found during a project-month}. We simply use $NewB_{im}$ to denote it. On \textsc{GitHub}, the issue can be of various types, e.g., discussion, new feature request, improvement request, and so on. To categorize these issues, software developers often employ some keywords to tag them. However, because tagging is often project-specific, we adopt Vasilescu et al.'s \cite{vasilescu2015quality} method to distinguish bug issues from other issue types in this study. We set up a list of bug-related keywords, including \textit{defect, error, bug, issue, mistake, incorrect, fault, and flaw}, and then search for these words
in both the issue tags and issue titles. If any tags or title of an issue contains at least one keyword, we identify it as a bug issue. Similarly, as the productivity data, we compute the number of new bug issues in every project-month. In addition to counting the newly found bugs in each project-month, our study also includes a second quality indicator: Monthly Bug Fix Rate ($BFR_{im}$), which is defined as: 
\begin{displaymath}
BFR_{im}=\frac{No.~of~Fixed~Bugs}{No.~of~Found~Bugs}, ~~for~project~i ~in~month~m.
\end{displaymath}
The Bug Fix Rate is one of the key metrics related to the defect removal process \cite{Distefano:2019:SSA:3351434.3338112,896249}. In fact, it partially represents the effectiveness of the quality assurance process by characterizing the birth (finding a new bug) to death (fixing a bug) process of defect removal \cite{levendel1990reliability}. If $BFR_{im} < 1$, it indicates that the project's quality risk is accumulating.


\subsection{Identifying the Elite Developers}

Following the method used in Hanisch et al.'s study~\cite{hanisch2018developers}, we leverage \textsc{GitHub}'s repository permission mechanism to identify the elite developer. Being an elite developer in a project means s/he obtained write permission for an organization's repository. By gaining this level of permission, the developer can perform many tasks on a repository without requesting, for example, directly pushing commits to a repository, creating and editing releases, and merging pull requests. In addition, with write permission of the repository, the developer is able to perform several types of administrative work, such as submitting code reviews that affect a pull request's mergeability, applying labels to tasks and milestones to the repository, and marking an issue as duplicate, which would let the issue lose public attention.

Unfortunately, \textsc{GitHub} does not allow anyone other than the repository owners to access the list of members obtaining specific permissions. We apply a permissions check mechanism to determine the elites. When a developer in the repository performs a task that requires the write permission, we tag this developer with ``elite-ship'' of the repository. 
As we observed in this study, we found that a project's elite developers might also suffer survival issue \cite{lin2017developer}, thus we set 90 days\footnote{Literature on survival analysis of open-source developer usually use 30 days or 90 days as a time window \cite{lin2017developer}. When we examined the raw data we have, we found that 30 days (1 month) is too short. But 90 days (3 months) is a good time interval to avoid rush decisions on judging if someone gains/loses the elite identity.} as the length of the ``elite-ship'', and use this time-window to filter developers who were inactive.
During this three-month period, if this developer performs any task that also requires the write permission, her ``elite-ship'' would get renewed for another three months, starting from the month when she performed the task.

Compared with other elite-developer-identification methods based on metrics or network \cite{Joblin:2017:CDC:3097368.3097389}, our methods have several advantages. First, our method takes a dynamic view of the status of being an elite developer. It is designated for the open-source community where developers have very high mobility in terms of entering and leaving\footnote{For company-sponsored projects, the mobility may also result from organizational and individual career changes.}. Secondly, our method reflects the socialization process of gaining power and status in a community. Thirdly, our method respects the fact that some developers may be nominated as elite developers before making substantial contributions, particularly in the company-sponsored projects. Lastly, our method avoids dealing with the marginal cases resulting from the arbitrarily set threshold, e.g., the 1/3 cut-off used in \cite{crowston2006core}.

\subsection{Data Analysis}
Tab. \ref{analysis_methods} presents the mapping between \textbf{RQs} and corresponding data-analysis methods. We will introduce them in detail in the rest of this section.
\begin{table}[!h]
    \centering
    \caption{Research questions and corresponding data analysis methods.}
    \begin{tabular}{ll}
    \toprule
    \textbf{RQs}& \textbf{Data Analysis Methods}\\
    \midrule
       $\mathbf{RQ_{1}}$  & Descriptive statistics \\
       $\mathbf{RQ_{2}}$  & Descriptive statistics, ANOVA\\
       $\mathbf{RQ_{3}}$ & Project-specific fixed effects Panel Regressions \\
       &(LSDV estimator with Diagnostics)\\
       \bottomrule
    \end{tabular}
    \label{analysis_methods}
\end{table}

All statistical analyses are performed with \textsc{R 3.4.1} \cite{R2008}, and its associated packages for macOS High Sierra (version 10.13.1). We follow the ASA's principles to present and interpret statistical significance \cite{doi:10.1080/00031305.2016.1154108}. 

\subsubsection{Summarize Activities for $\mathbf{RQ_{1}}$}
Answering $\mathbf{RQ_{1}}$ does not require complicated analysis techniques. We use descriptive statistics to derive results and findings for this research question. Note that we code the raw \textsc{GitHub} activities into four broad activity categories (communicative, organizational, supportive, and typical) according to \cite{sonnentag1995excellent} (described in Section~\ref{sec:datapreparation}). Doing so helps us to derive meaningful insights instead of fragile, overly detailed information in the raw activities. For all sampled projects, we calculate the total of elite developers' activities over the four broad categories. Thus, we have a 4-tuple for each project as follows:
\begin{displaymath}
<Com, Org, Sup, Typ>
\end{displaymath}
We also compute the percentage of elite developers' activities over the entire project's activities. All results are reported in Section~\ref{sec:rq1results}.

\subsubsection{Identifying Activity Trend for $\mathbf{RQ_{2}}$}\label{sssec:rq2analysis}
To answer $\mathbf{RQ_{2}}$, we first group the activities according to the month of their occurrences. Then, similarly, for a project \emph{i} in each month {m}, we can calculate a similar 4-tuple:
\begin{displaymath}
<Com_{im}, Org_{im}, Sup_{im}, Typ_{im}>
\end{displaymath}
where $i \in \{1,..., i,..., 20\}$, and $m \in \{1,...,m,...,36\}$.

Since the different projects have different numbers of elite developers, cross-project comparisons require to average the project-level data to individual-level. We simply calculate the average activities per developer over the four categories. Then, we can calculate the individualized monthly growth rates of activities in each category for each project. Given that there are 20 projects, for each category, we have 20 growth rates. We use one-way ANOVA to see if there is any difference across the four categories regarding the growth rates.  

\subsubsection{Identifying Correlations with Project Outcomes for $\mathbf{RQ_{3}}$}
Answering $\mathbf{RQ_{1}}$ and $\mathbf{RQ_{2}}$ provides the data we need to answer $\mathbf{RQ_{3}}$. Before discussing the analysis methods, we first examine the data.

We want to investigate the correlations between a project's elite developers' effort distributions and project outcomes. The \emph{independent variables} are the effort distributions over the four categories of activities, which can be easily extracted from the collected data. The dependent variables are four indicators of project outcomes (productivity: $NewC_{im}$, $BCT_{im}$; quality: $NewB_{im}$, $BFR_{im}$), which are adapted from the prior software engineering literature. Given that we have broken a project's data into months when answering $\mathbf{RQ_{2}}$ and using ``month'' as the analysis unit, we have one data case for each project $i$ at each month $m$. Therefore, we have 720 (20 projects $\times$ 36 months) data cases, in total. Each data case is in the following form:
\begin{displaymath}
<NewC_{im},~BCT_{im},~NewB_{im},~BFR_{im},~\overline{S-Com_{im}},~ \overline{S-Org_{im}},~\overline{S-Sup_{im}},~\overline{S-Typ_{im}}>
\end{displaymath}
where $i \in \{1,..., i,..., 20\}$, and $m \in \{1,...,m,...,36\}$.

The $\overline{S-Com_{im}}$ represents the share of communicative activities in all four categories of activities per elite developer for project $i$ in month $m$. Similar denotations apply to the other three. Note that,
\begin{equation}
\overline{S-Com_{im}}+\overline{S-Org_{im}}+\overline{S-Sup_{im}}+\overline{S-Typ_{im}} =1 
\label{sum}
\end{equation}

\noindent Answering $\mathbf{RQ_{3}}$ is identifying the relationships between these four independent variables and four dependent variables $NewC_{im}$, $BugC_{im}$, $NewB_{im}$, and $BFR_{im}$. A natural solution is performing regression analysis. Our data is panel data (cross-sectional: from 20 projects; longitudinal: 36 months per project). Thus, simple OLS multivariate linear regression is not a proper technique because we cannot assume there is no difference among the 20 projects and 36 data points.

To correctly identify the relationships, we employ Econometric methods to deal with the panel data \cite{wooldridge2015introductory}. Intuitively, each project has its own characteristics, so we use the project-specific fixed effects models\footnote{We also empirically perform model diagnostics which proves fixed-effect models are better than both OLS and random effects models, see Section 4.3.1.}. The analyses actually estimate parameters for the following four regression equations (2) to (5).
\begin{equation}
   NewC_{im} = \beta_1 \times \overline{S-Com_{im}}  +\beta_2 \times \overline{S-Org_{im}} + \beta_3 \times \overline{S-Sup_{im}} +\alpha_i + u_{it}
\label{re_commit}
\end{equation}

\begin{equation}
\begin{split}
   BCT_{im} &= \beta_1 \times \overline{S-Com_{im}}  +\beta_2 \times \overline{S-Org_{im}} + \beta_3 \times \overline{S-Sup_{im}} +\alpha_i + u_{it}
\end{split}
\label{re_issue}
\end{equation}

\begin{equation}
\begin{split}
   NewB_{im} &= \beta_1 \times \overline{S-Com_{im}}  +\beta_2 \times \overline{S-Org_{im}} + \beta_3 \times \overline{S-Sup_{im}} +\alpha_i + u_{it}
\end{split}
\label{re_issue}
\end{equation}

\begin{equation}
\begin{split}
   BFR_{im} &= \beta_1 \times \overline{S-Com_{im}}  +\beta_2 \times \overline{S-Org_{im}} + \beta_3 \times \overline{S-Sup_{im}} +\alpha_i + u_{it}
\end{split}
\label{re_issue}
\end{equation}

Note that we do not include $\overline{S-Typ_{im}}$ into Regression Equations \ref{re_commit}--\ref{re_issue}. The reason is straightforward: the sum of $\overline{S-Typ_{im}}$ and the other three is always ``1'' according to Eq. \ref{sum}. Thus, it is perfectly correlated with the other three. Including it will lead to a significant multicollinearity problem\footnote{In fact, no coefficient can be estimated for it in \texttt{R}.}.

For each dependent variable, we use the least-squares dummy variables (LSDV) estimator to estimate the parameters in the project-specific fixed effects models. After we finish the model estimation, we perform a series of regression diagnostics for examining the time-specific effects and empirically justifying the use of fixed effects models. These regression diagnostics include: time-fixed effects testing, F-test for (\texttt{pFtest}), Hausman Test (\texttt{pHtest}), Heteroskedasticity testing, and so on. Given that our sampled projects consist of 11 company-sponsored projects and 9 non-company-sponsored ones. It is natural to investigate if effort distributions' impacts on project outcomes are sensitive to these project characteristics. Therefore, we perform the same regression analyses to the two sub-samples. The results are reported accordingly. All the panel regressions, if not otherwise stated, are performed with \texttt{R}'s \texttt{plm} package \cite{plm2008}.

\section{Results and Findings}

In this section, we report the results and findings. We organize them according to the three $\mathbf{RQ}$s. All data has been made publicly available for download\footnote{All experiment result, including intermediate outputs and raw data, can be downloaded at: \url{https://drive.google.com/drive/folders/10ibmz2svPRf3jfRtm7mbiouo9ATaYAoB}}.

\begin{table}[!h]
    \small
    \centering
    \caption{Activities amount that has happened on each sampled project.}
    \begin{tabular}{lllllllll}
    \toprule
        \multirow{2}{*}{\textbf{Project}} &   \multicolumn{2}{c}{\textbf{Com.}}    &   \multicolumn{2}{c}{\textbf{Org.}}    &   \multicolumn{2}{c}{\textbf{Sup.}}    &   \multicolumn{2}{c}{\textbf{Typ.}} \\ 
        \cmidrule(lr){2-3}
        \cmidrule(lr){4-5}
        \cmidrule(lr){6-7}
        \cmidrule(lr){8-9}
        & Total & Elite\% & Total & Elite\% & Total & Elite\% & Total & Elite\% \\
    \midrule
        \textsf{Aframe}          & 4908  & \Chart{0.46}  & 455   & \Chart{0.99}  & 19400 & \Chart{0.85}  & 5180  & \Chart{0.72} \\
        \textsf{Alamofire}       & 5967  & \Chart{0.18}  & 1773  & \Chart{1.00}  & 11906 & \Chart{0.61}  & 1465  & \Chart{0.62} \\
        \textsf{Exoplayer}       & 9293  & \Chart{0.32}  & 2293  & \Chart{1.00}  & 22197 & \Chart{0.71}  & 5361  & \Chart{0.87} \\
        \textsf{finagle}         & 2488  & \Chart{0.30}  & 46    & \Chart{0.93}  & 2947  & \Chart{0.46}  & 2753  & \Chart{0.49} \\
        \textsf{fresco}          & 5283  & \Chart{0.29}  & 290   & \Chart{1.00}  & 10481 & \Chart{0.64}  & 1923  & \Chart{0.77} \\
        \textsf{guava}           & 3161  & \Chart{0.29}  & 664   & \Chart{0.99}  & 7724  & \Chart{0.71}  & 2239  & \Chart{0.53} \\
        \textsf{immutable-js}    & 2909  & \Chart{0.15}  & 28    & \Chart{0.75}  & 5869  & \Chart{0.59}  & 1057  & \Chart{0.52} \\
        \textsf{jest}            & 19073 & \Chart{0.36}  & 1025  & \Chart{0.99}  & 39995 & \Chart{0.66}  & 5015  & \Chart{0.43} \\
        \textsf{marko}           & 1937  & \Chart{0.38}  & 403   & \Chart{0.95}  & 5525  & \Chart{0.79}  & 2956  & \Chart{0.93} \\
        \textsf{Moya}            & 5376  & \Chart{0.43}  & 416   & \Chart{0.61}  & 22808 & \Chart{0.42}  & 2860  & \Chart{0.73} \\
        \textsf{nightmare}       & 3105  & \Chart{0.14}  & 24    & \Chart{1.00}  & 4963  & \Chart{0.48}  & 892   & \Chart{0.50} \\
        \textsf{rclone}          & 7475  & \Chart{0.23}  & 182   & \Chart{1.00}  & 15243 & \Chart{0.66}  & 2781  & \Chart{0.80} \\
        \textsf{react}           & 35086 & \Chart{0.37}  & 3730  & \Chart{1.00}  & 83036 & \Chart{0.74}  & 9640  & \Chart{0.59} \\
        \textsf{recharts}        & 3199  & \Chart{0.15}  & 54    & \Chart{0.85}  & 4980  & \Chart{0.41}  & 1396  & \Chart{0.66} \\
        \textsf{splitebrowser}   & 6129  & \Chart{0.42}  & 493   & \Chart{1.00}  & 11589 & \Chart{0.70}  & 1751  & \Chart{0.84} \\
        \textsf{stf}             & 1694  & \Chart{0.33}  & 25    & \Chart{0.64}  & 2672  & \Chart{0.55}  & 837   & \Chart{0.69} \\
        \textsf{tensorflow}      & 97940 & \Chart{0.48}  & 28236 & \Chart{0.92}  & 183485& \Chart{0.75}  & 43029 & \Chart{0.50} \\
        \textsf{tesseract}       & 4870  & \Chart{0.35}  & 116   & \Chart{1.00}  & 9805  & \Chart{0.59}  & 2512  & \Chart{0.55} \\
        \textsf{tidb}            & 16240 & \Chart{0.80}  & 1944  & \Chart{0.98}  & 45451 & \Chart{0.91}  & 8305  & \Chart{0.89} \\
        \textsf{ZeroNet}         & 2853  & \Chart{0.27}  & 120   & \Chart{1.00}  & 4881  & \Chart{0.56}  & 2650  & \Chart{0.79} \\
    \midrule
    \textbf{Mean}       & \textbf{11949.30}    & \Chart{0.34} & \textbf{2115.85} & \Chart{0.93} & \textbf{25747.85} & \Chart{0.64} & \textbf{5230.10} & \Chart{0.67} \\
    \bottomrule
    \end{tabular}
    \label{statisticsAC}
\end{table}

\subsection{$\mathbf{RQ_1}$: Elite Developers' Activities.}
\label{sec:rq1results}

Tab. \ref{statisticsAC} provides the basic demographic statistics of the activities in each project according to their categories. Except for the communicative activities, elite developers perform over 50\% of the activities for those in all three of the remaining categories. For each project, our results have confirmed the finding from other studies on the core or elite developers of Open Source communities, e.g., \cite{Mockus:2002:TCS:567793.567795, ducheneaut2005socialization, wagstrom2012roles}, and elite developers in the community contributed most of the source-code submission. In our sample, 67 percent of typical development tasks are performed by a project's elite developers.

\begin{wrapfigure}{R}{0.35\textwidth}
\centering
\includegraphics[width = 0.32\columnwidth]{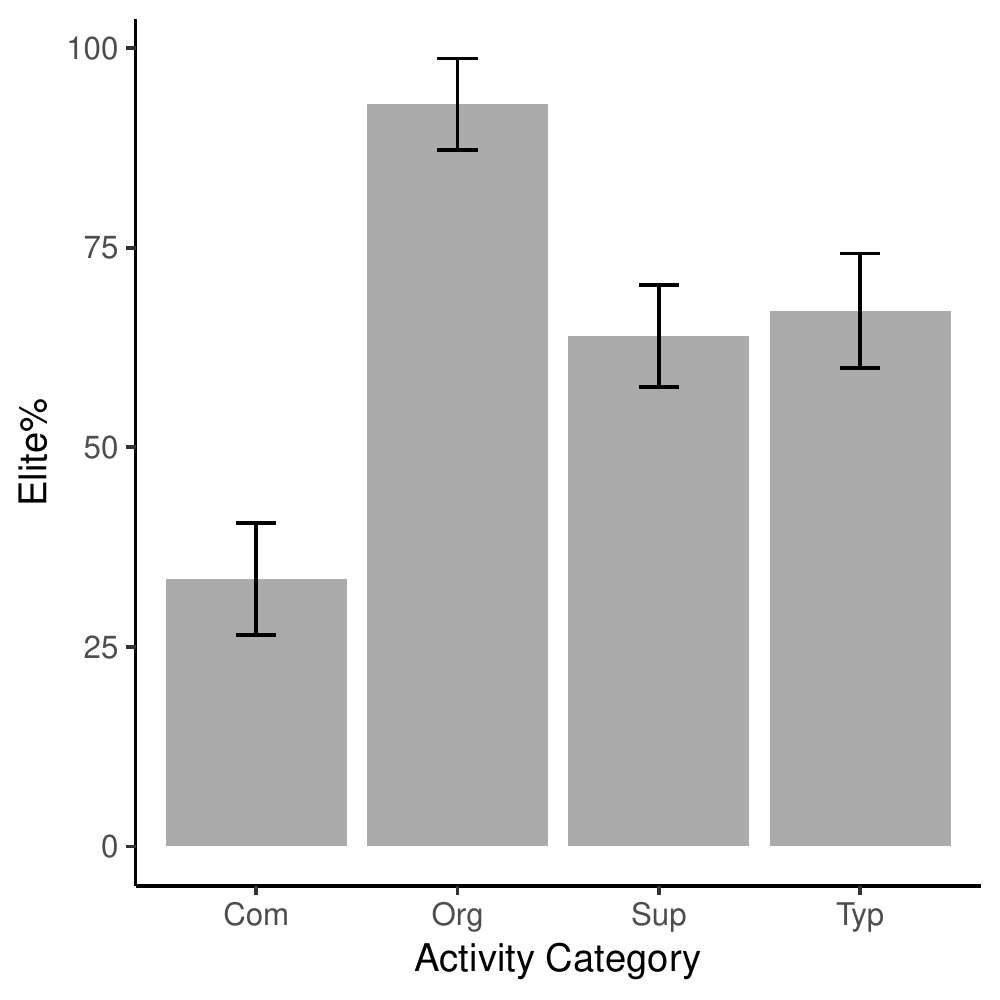}
\caption{The distributions of elite developers' activity shares in each activity category over 20 projects.}
\vspace{-1em}
\label{distribution}
\end{wrapfigure}

In addition to elites' code submission, we also found empirical evidence that elite developers are also ``responsible'' for most other types of events. Besides organizational events (according to our definitions, most organizational events automatically require the write permission), elite developers perform over 60\% of supportive activities and even created 34\% of communicative activities. See Fig. \ref{distribution} for the percentage distribution of elites' contribution.

\begin{figure}[h]
\centering
\includegraphics[width = 0.70\columnwidth]{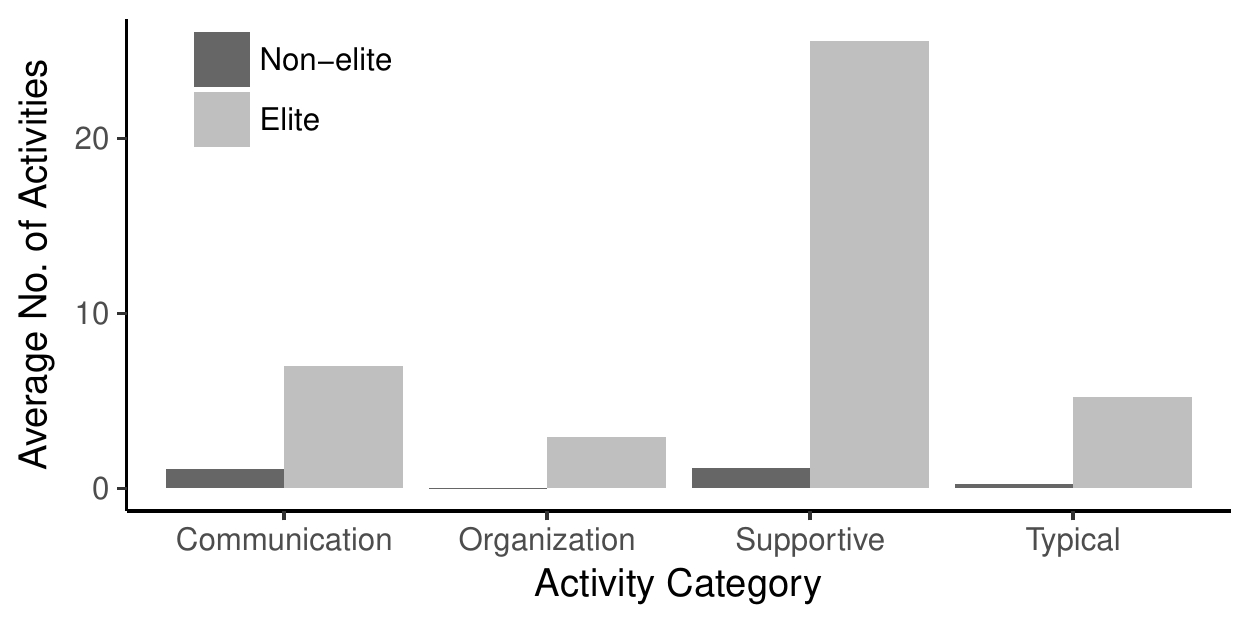}
\caption{Average monthly activities comparisons between elite and non-elite developers.}
\label{comparison}
\end{figure}

Moreover, comparing to non-elite developers, the average numbers of activities performed by an elite developer per month are much higher in all categories (see Fig. \ref{comparison}). We observe orders of magnitude differences between them. On average, an elite developer performs 7 times more communication activities, 145 times more organizational activities, 22 times more supportive activities, and 22 times more typical activities than a non-elite developer per month. Thus, on an individual basis, we argue that elite developers may have major impacts on projects based on their activity amount.


\noindent{\textbf{Answers to $\mathbf{RQ_1}$.}}
Based on the events in each category, we can answer $\mathbf{RQ_1}$ as follows:

\begin{framed}
\noindent \textit{On \textsc{GitHub}, elite developers have contributed to the project in various ways in addition to performing over 60\% code contributions. They need to manage the community by delegating tasks to other developers with special expertise, managing parallel development among contributors, creating documentations for the project, and also participating in discussions with teammates, external developers, and peripheral users.}
\end{framed}

\subsection{$\mathbf{RQ_2}$: The Evolution of Elite Developers' Activities.}
\begin{figure}[!h]
\centering
\includegraphics[width = 0.9\columnwidth]{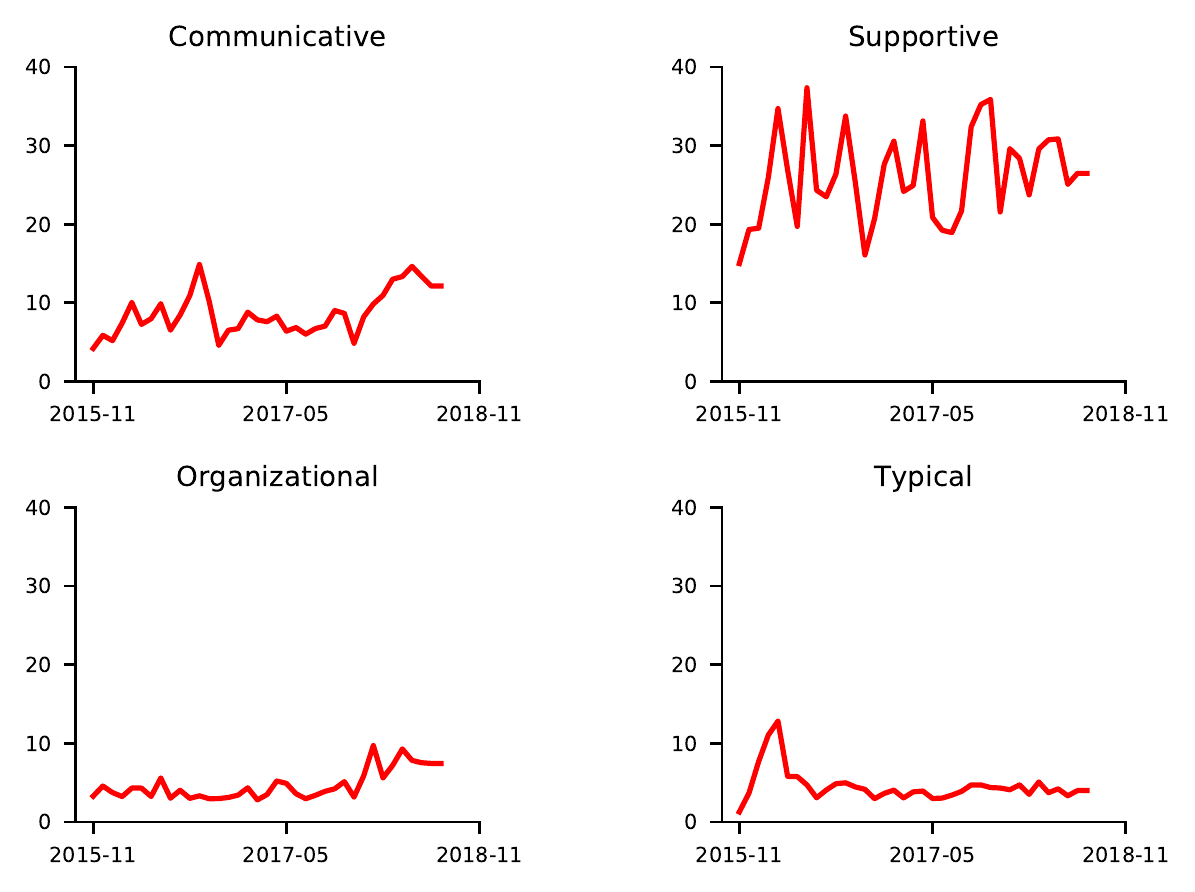}
\caption{Trends of individual elite developer's activities in the four activity categories of the \textit{Tensorflow}.}
\label{sample}
\end{figure}

\subsubsection{Individual Activities of Elite.} For the most complex project in our project sample, Tensorflow, we found that there is a steady increase in communicative, supportive, and organizational events for each elite developer (shown in Fig. \ref{sample}). Though supportive events change dramatically because of the period of software patches and releases, it still shows an increase in the longitudinal perspective. The increase of organizational events may be due to the scale increase of the team (the number of active elite developers has increased from 29 to 270 for \textit{Tensorflow}). However, we found the amount of code submissions by elite developers has stabilized since the initial project release phase, even for fast growing projects such as Tensorflow. In order to verify whether this focus shifts of elite developers are common in our sampled projects, we test the differences in growth rates of activity categories as the next.
\subsubsection{Comparing Growth Rates of the Four Types of Activities}
As mentioned in Section \ref{sssec:rq2analysis},
we calculate the average monthly growth rates of activities per elite developer over the communicative, supportive, and typical activities\footnote{For organizational activities, many months do not record such type of activities. This prohibits us from calculating the growth rate.} for each project. Thus, we have 20 growth rates for these three categories of activities. We then perform one-way ANOVA to test if there is any difference in growth rates.

The results shows significant differences ($F_{(2, 57)} = 8.452$, $p<0.001$). We perform the post-hoc analysis using the Tukey's HSD test to identify the differences between the three categories. The results indicate the growth rates of typical activities are significantly lower than the growth rates of the other two (Typical vs. Communicative: $p=0.002$, Typical vs. Supportive: $p=0.002$). In fact, elite developers' typical activities even decrease over the time (average growth rate = $-1.63$\%). Though this number seems not that big, it actually means an elite developer only does half of the technical work she used to do 3 years ago. Meanwhile, their work on communicative and supportive are doubled in the same period. 

We do not perform the same ANOVA procedures to the non-elite's data for cross-group comparisons (i.e., elite vs. non-elite) due to practical constraints. In many months, the non-elite's activity counts are $0$. Thus, calculating growth rates would lead to many ``division by zero'' problems. However, qualitatively, we could not observe any significant increases on the three types of non-technical activities over the time, while the numbers of non-elite's technical activities in each project-month do increase over the time.

\noindent
\textbf{Answers to $\mathbf{RQ_2}$.} Based on the result of one-way ANOVA test and Tukey's HSD test, we can answer $\mathbf{RQ_2}$ as follows:

\begin{framed}
\noindent \textit{With the progress of the project, an elite developer tends to put more efforts into communicative and supportive activities while she significantly reduces her involvements in typical development activities.}
\end{framed}

\begin{sidewaystable}
    \centering
    \begin{threeparttable}
     \caption{Regression models for project productivity.\label{tab:regp}}
    \begin{tabular}{ldddddd}
    \toprule
   & \multicolumn{2}{c}{Whole Sample} & \multicolumn{2}{c}{Sub-sample (Non-Company)} &\multicolumn{2}{c}{Sub-sample (Company)}\\
   &\multicolumn{1}{c}{New Commit}&\multicolumn{1}{c}{Bug Cycle Time}&\multicolumn{1}{c}{New Commit}&\multicolumn{1}{c}{Bug Cycle Time}&\multicolumn{1}{c}{New Commit}&\multicolumn{1}{c}{Bug Cycle Time} \\
         &\multicolumn{1}{c}{Model P1 ($\beta$)}&\multicolumn{1}{c}{Model P2 ($\beta$)}&\multicolumn{1}{c}{Model P3 ($\beta$)}&\multicolumn{1}{c}{Model P4 ($\beta$)}&\multicolumn{1}{c}{Model P5 ($\beta$)}&\multicolumn{1}{c}{Model P6 ($\beta$)}  \\
         &(\emph{SE})&(\emph{SE})&(\emph{SE})&(\emph{SE})&(\emph{SE})&(\emph{SE}) \\
    \midrule
         $\overline{S-Com_{im}}$ &-155.96$**$ &-170.71&-125.07$***$&-381.80&-228.68$*$&-16.44\\
         &(49.20)&(143.96)&(28.11)&(212.09)&(92.85)&(202.66)\\
         $\overline{S-Org_{im}}$ &1.38&-148.61$*$&-137.28$*$&-214.71$*$&203.10&-153.85\\
         &(121.25)&(54.65)&(70.24)&(129.6)&(223.47)&(187.49)\\
         $\overline{S-Sup_{im}}$ &-138.21$***$&473.60$***$&-91.25$***$&553.17$**$&-204.66$**$&401.30$**$\\
         &(35.52)&(103.94)&(20.34)&(153.39)&(66.16)&(144.40)\\
         \emph{Unobserved time-invariant effects} $(\alpha_{i}) ^{\P}$ &-.-$***$&-.-$***$&-.-$***$&-.-$***$&-.-$***$&-.-$***$\\
         \midrule
         \emph{Multiple} $ R^2$&0.884&0.745&0.686&0.740&0.891&0.751\\
         \emph{Adjusted Multiple} $R^2$&0.880&0.736&0.673&0.730&0.887&0.742\\
         $F^{\dagger}$& 231.10$***$&88.35$***$&60.45$***$&73.89&222.00$***$&82.44$***$\\
         \bottomrule
         
    \end{tabular}
   \begin{tablenotes}
   \item *: $p < 0.05$,  **: $p < 0.01$,  ***: $p < 0.001$.
   \item $^{\P}$: The unobserved time-invariant effects are a vector rather than a single coefficient. To keep the paper concise, we do not include them, instead, we show the significant levels of them. 
   \item $^{\dagger}$: For Model 1 -- 2: degrees of freedom ($DF$s) are (23, 697); for Model 3 -- 4:  degrees of freedom are (12, 312); for Model 5 -- 6: degrees of freedom are (14, 382).
   \end{tablenotes}
    \end{threeparttable}
\end{sidewaystable}

\subsection{$\mathbf{RQ_3}$: Elite Developers' Activities' Impacts on Project Outcomes.}

We present our findings to $\mathbf{RQ_3}$ with a series regression models (Tab. \ref{tab:regp} and \ref{tab:regq}) characterizing relationships between the shares of activities in the three categories and the product productivity and quality indicators. These models are developed using the econometrics techniques discussed in Section 3.6.3. We also perform regression diagnostics to empirically examine the justification of using fixed effects models in model development. For all 12 regressions, fixed effects models are better choices than pooled OLS and random models.

Note that all regression models establish \emph{correlations} only, rather than \emph{causalities}. However, when interpreting the results, we may give some propositions implying possible but not definitive causalities, which is a common practice in data-driven research related to human and social factors \cite{boyd2012, Cowls2015, Einav2014}. All these implied causalities shall not be considered as established without further confirmatory studies \cite{savage2007}.


\subsubsection{Regression Results of Project Productivity}

Tab. \ref{tab:regp} summarizes the results of the regression models for the two project productivity indicators: the no. of new commit of project $i$ in month $m$ ($NewC_{im}$), and the average bug cycle time of project $i$ in month $m$ ($BCT_{im}$). Models P1 and P2 use the data of all 20 sampled projects, thus represent whole sample regression results. Models P3 and P4 use the data of 9 non-company-sponsored projects, while models P5 and P6 use the data of 11 company-sponsored projects. Thus, models P3--P6 are representing sub-sample regression results. Now let us have a look at what these models indicate. 

\noindent \underline{\emph{A. Project Productivity---Whole Sample Regression Results}} 

\noindent In Model P1, two independent variables ($\overline{S-Com_{im}}$, $\overline{S-Sup_{im}}$) are significant; and both have negative regression coefficients ($-155.96$, $-138.21$). 
This implies that \textbf{negative correlations} between the effort elite developers put on communicative and supportive activities, and the no. of new commits in each project-month. A possible interpretation of the results is as follows. We already know that elite developers are still major contributors of the source code. When they invest more efforts on non-technical activities such as communicative and supportive ones, they may have less time to contribute to the source code; thus, the whole project may have fewer new commits (productivity loss). 

In Model P2 which the dependent variable is $BCT_{im}$, two independent variables ($\overline{S-Org_{im}}$, $\overline{S-Sup_{im}}$) are significant. $\overline{S-Org_{im}}$ has a negative regression coefficient ($-148.61$), while $\overline{S-Sup_{im}}$ has a positive coefficient ($473.60$). 
Obviously, there are \textbf{negative correlations} between elite developers' efforts in organizational activities and average bug cycle time in each project-month, and \textbf{positive correlations} between elite developers' efforts in supportive activities and average bug cycle time in each project-month. 
Given that the activities of managing bug fix and code review are in the ``organizational'' category (see Fig. 2), more elites' efforts in activities in this category might help to shorten the bug cycle time. Meanwhile, similar to the results and interpretation for Model P1, performing more supportive activities may occupy elite developers' time on fixing bugs, and thus lead to longer bug cycle time (productivity loss). Since $\overline{S-Sup_{im}}$'s effect is much stronger than $\overline{S-Org_{im}}$'s and its shares are often much more than $\overline{S-Org_{im}}$'s (Avg.: 0.61 vs. 0.03), we could expect an overall effect of longer bug cycle time (productivity loss).

\noindent \underline{\emph{B. Project Productivity---Sub-Sample Regression Results.}} 

\noindent The regression results in Models P3--P6 are pretty much similar to those in Models P1 and P2 with some minor differences. 
Let us first have a look at the regression models based on non-company-sponsor projects' data (Models P3 and P4).
In Model P3, $\overline{S-Org_{im}}$ becomes a significant variable, indicating that performing more organizational activities is also \textbf{negatively correlated} with the no. of new commits in each project-month (productivity loss). 
In Model P4, the correlations between efforts on each category and the average bug cycle time in each project-month ($BCT_{im}$) are the same. 
For the regression models based on company-sponsor projects' data (Models P5 and P6), correlations in Model P5 are as same as those in Model P1. However, in Model P6, $\overline{S-Org_{im}}$ is no longer significant. A possible explanation may be that: company-sponsored projects often have established routine bug fixing processes, and hence elite developers' mediation in this process is not as important.

In addition, the adjusted $R^2$s of Models P5 \& P6 are higher than Models P3 \& P4. Particularly, Model P5's is over 20\% higher than Model P3's. 
These differences indicate that models built around the elite developers' activities work better for company-sponsored projects.

\subsubsection{Regression Results of Project Quality}
We have briefly discussed the regression results of project productivity. Now, let us turn to the regression results of project quality. Tab. \ref{tab:regq} summarizes the results of the regression models for the two project productivity indicators: the no. of new bugs of project $i$ in month $m$ ($NewC_{im}$), and the big fixed rate of project $i$ in month $m$ ($BFR_{im}$). Similarly, Models Q1 and Q2 use the data of all 20 sampled projects, thus represent whole sample regression results. Models Q3 and Q4 use the data of 9 non-company-sponsored projects, while models Q5 and Q6 use the data of 11 company-sponsored projects. Thus, models Q3--Q6 are representing sub-sample regression results.

\begin{sidewaystable}
    \centering
    \begin{threeparttable}
     \caption{Regression models for project quality.\label{tab:regq}}
    \begin{tabular}{ldddddd}
    \toprule
   & \multicolumn{2}{c}{Whole Sample} & \multicolumn{2}{c}{Sub-sample (Non-Company)} &\multicolumn{2}{c}{Sub-sample (Company)}\\
   &\multicolumn{1}{c}{New Bug}&\multicolumn{1}{c}{Bug Fix Rate}&\multicolumn{1}{c}{New Bug}&\multicolumn{1}{c}{Bug Fix Rate}&\multicolumn{1}{c}{New Bug}&\multicolumn{1}{c}{Bug Fix Rate} \\
         &\multicolumn{1}{c}{Model Q1
         ($\beta$)}&\multicolumn{1}{c}{Model Q2 ($\beta$)}&\multicolumn{1}{c}{Model Q3 ($\beta$)}&\multicolumn{1}{c}{Model Q4 ($\beta$)}&\multicolumn{1}{c}{Model Q5 ($\beta$)}&\multicolumn{1}{c}{Model Q6 ($\beta$)}  \\
         &(\emph{SE})&(\emph{SE})&(\emph{SE})&(\emph{SE})&(\emph{SE})&(\emph{SE}) \\
    \midrule
         $\overline{S-Com_{im}}$ &4.13 &-2.24$***$&1.43&-1.28$***$&5.23&-3.11$***$\\
         &(5.26)&(0.35)&(3.91)&(0.34)&(9.59)&(0.61)\\
         $\overline{S-Org_{im}}$ &50.13$***$&-0.63&6.48&-0.62&88.69$***$&-0.35\\
         &(12.97)&(0.87)&(9.77)&(0.85)&(23.09)&(1.47)\\
         $\overline{S-Sup_{im}}$ &18.31$***$&1.12$***$&-7.67$**$&0.60$*$&26.99$***$&1.50$***$\\
         &(3.80)&(0.26)&(2.83)&(0.25)&(6.84)&(0.44)\\
         \emph{Unobserved time-invariant effects} $(\alpha_{i}) ^{\P}$ &-.-$***$&-.-$***$&-.-$***$&-.-$***$&-.-$***$&-.-$***$\\
         \midrule
         \emph{Multiple} $ R^2$&0.857&0.649&0.667&0.761&0.871&0.608\\
         \emph{Adjusted Multiple} $R^2$&0.853&0.637&0.654&0.752&0.866&0.594\\
         $F^{\dagger}$& 186.2$***$&56.09$***$&52.05$***$&83.07$***$&184.2$***$&42.31$***$\\
         \bottomrule
         
    \end{tabular}
   \begin{tablenotes}
   \item *: $p < 0.05$,  **: $p < 0.01$,  ***: $p < 0.001$.
   \item $^{\P}$: The unobserved time-invariant effects are a vector rather than a single coefficient. To keep the paper concise, we do not include them, instead, we show the significant levels of them. 
   \item $^{\dagger}$: For Model 1 -- 2: degrees of freedom ($DF$s) are (23, 697); for Model 3 -- 4:  degrees of freedom are (12, 312); for Model 5 -- 6: degrees of freedom are (14, 382).
   \end{tablenotes}
    \end{threeparttable}
\end{sidewaystable}

\noindent \underline{\emph{A. Project Quality---Whole Sample Regression Results}}

In Model Q1, the quality indicator is $NewC_{im}$. There are two significant independent variables ($\overline{S-Org_{im}}$, $\overline{S-Sup_{im}}$); and both have positive regression coefficients (50.13, 18.31). This indicates \textbf{positive correlations} between the effort elite developers put on organizational and supportive activities, and the no. of new bugs found in each project-month. The interpretation of the results shall be similar to the above. Doing non-technical work may make the elite have less time to work on code. Thus, non-elite developers may have to take more responsibilities on source code development. Their code may contain more bug  (quality loss).

In Model Q2, the quality indicator is $BFR_{im}$. 
Two independent variables are significant ($\overline{S-Com_{im}}$, $\overline{S-Sup_{im}}$). 
$\overline{S-Com_{im}}$'s coefficient is negative, signifying \textbf{negative correlations} between the effort elite developers put on communicative activities and each month's bug fix rate. 
Meanwhile, $\overline{S-Sup_{im}}$'s coefficient is positive, indicating \textbf{positive correlations} between the elite's efforts in supportive activities and each month's bug fix rate. Interpreting such correlations may be a bit tricky. 
For the negative correlations between $\overline{S-Com_{im}}$ and $BFR_{im}$, we can interpret it in a way similar to the previous ones. The positive correlations between $\overline{S-Sup_{im}}$ and $BFR_{im}$ may suggest that: by putting more efforts into supportive activities, elite developers help to make the defect removal process work well\footnote{Recall that $BFR_{im}$ is indeed a quality process metrics, see Section{}.}.  
Since $\overline{S-Com_{im}}$'s share are only about 1/4 of $\overline{S-Sup_{im}}$'s (Avg.: 0.16 vs. 0.61) and its negative coefficient is just twice of $\overline{S-Sup_{im}}$'s (-2.24 vs. 1.12), we could expect an overall effect of higher bug fix rate in average (quality gain).

\begin{figure}[!h]
\centering
\subfigure[New Commit]{%
\label{fig:first}%
\includegraphics[width = 0.45\columnwidth]{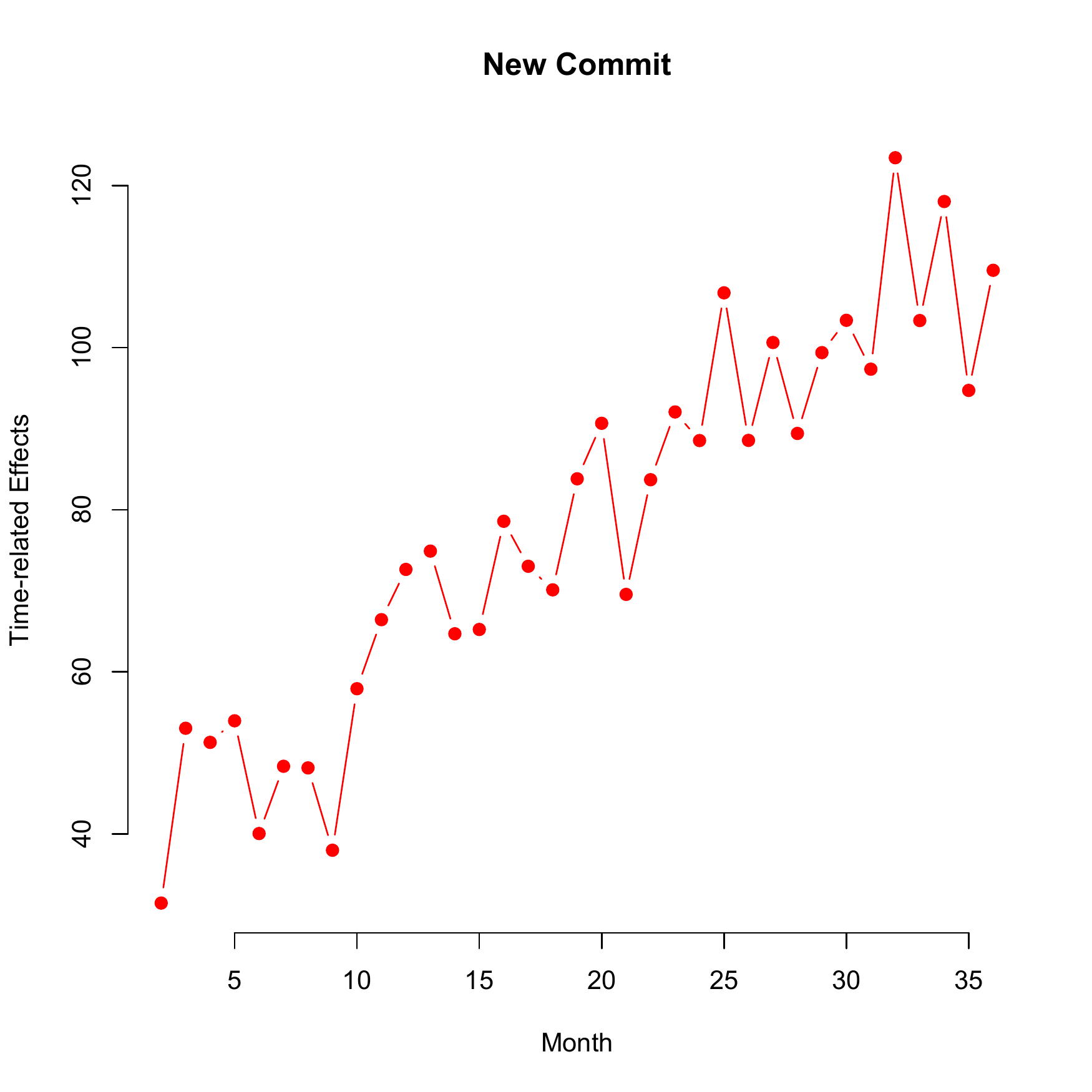}}%
\qquad
\subfigure[Bug Cycle Time]{%
\label{fig:second}%
\includegraphics[width = 0.45\columnwidth]{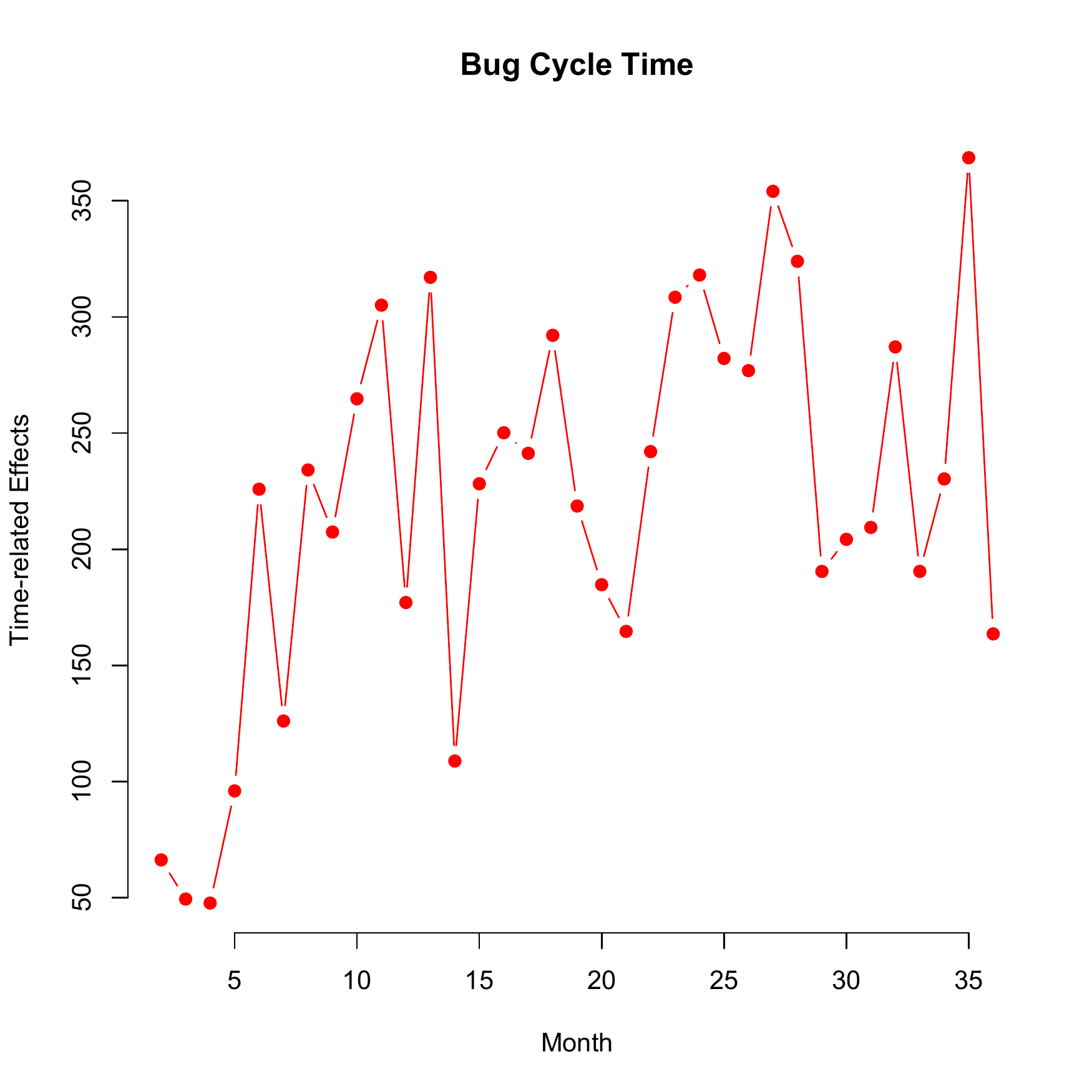}}%
\\
\subfigure[New Bug]{%
\label{fig:third}%
\includegraphics[width = 0.45\columnwidth]{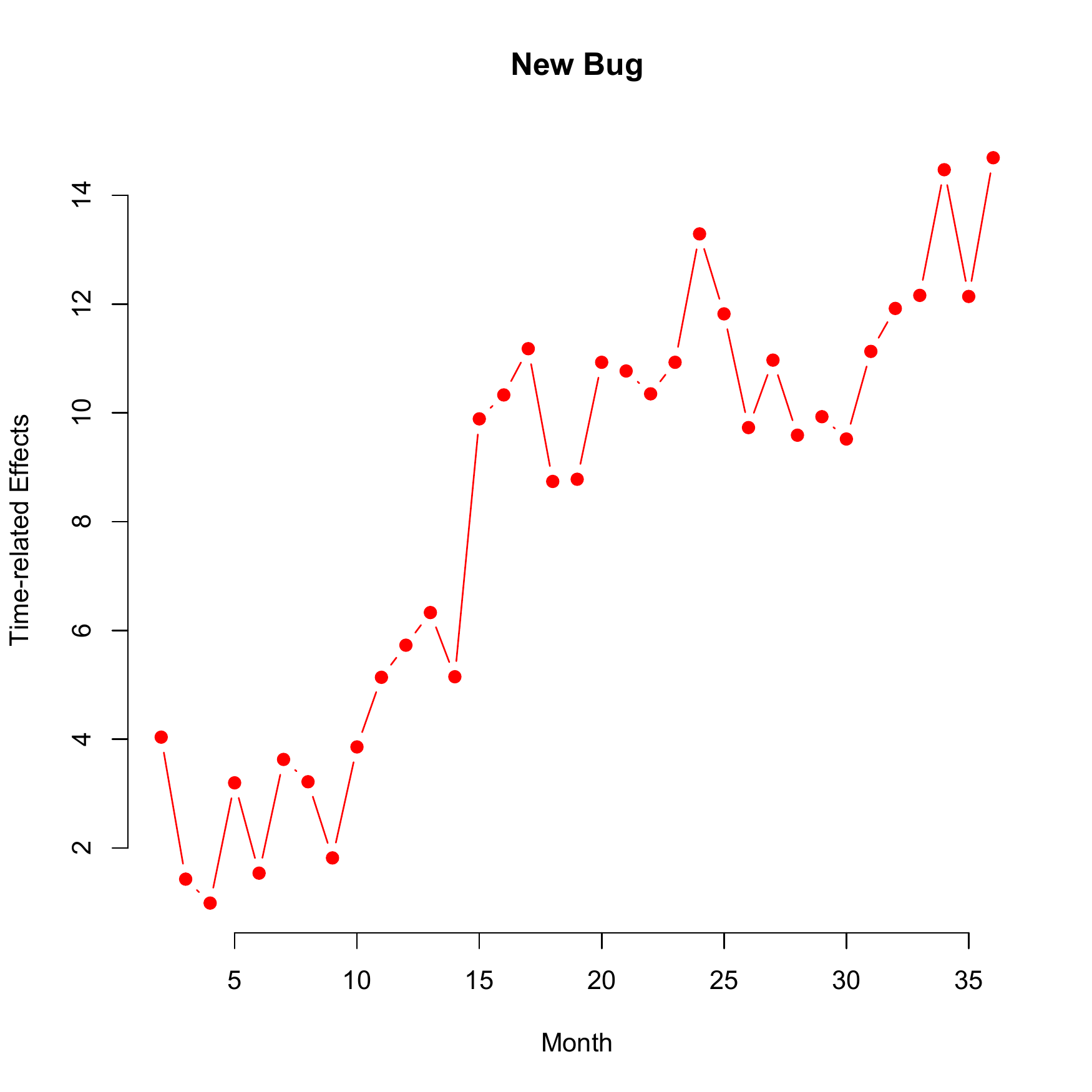}}%
\qquad
\subfigure[Bug Fix Rate]{%
\label{fig:fourth}%
\includegraphics[width = 0.45\columnwidth]{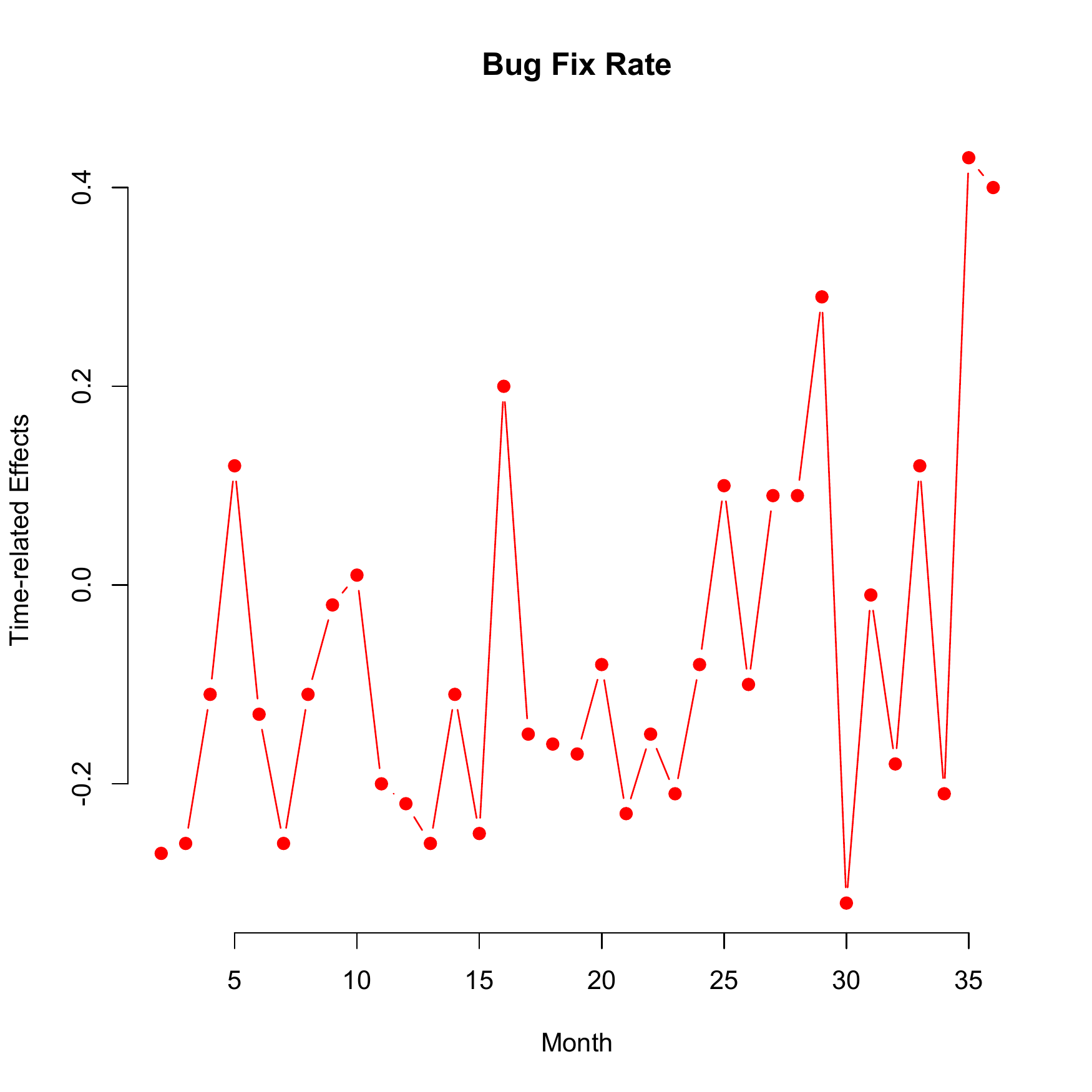}}%
\caption{Changes of time-related effects\label{fig:time}.}
\end{figure}
\noindent \underline{\emph{B. Project Quality---Sub-Sample Regression Results}} 

Again, we split the whole sample dataset into two sub-sample datasets according to whether a project is sponsored by a commercial company, and develop regression models using them. Models Q3 and Q4 are based on non-company-sponsored projects' data. In Model Q3, only $\overline{S-Sup_{im}}$ is significant. Efforts on organizational activities are not negatively correlated with the no. of new bugs in each project month. Model Q4 is similar to Model Q2. In general, the effects in Models Q5 \& Q6 are quite similar to those in Models Q1 \& Q2.

\subsubsection{Time-related Effects}

To further explore the time-related effects, we perform time-fixed effects testing for the four whole sample models (Models P1, P2 in Tab. \ref{tab:regp}; and Models Q1, Q2 in Tab. \ref{tab:regq}). 

For Model P1, where the number of new commits in each project-month is the dependent variable, the time-fixed effects model is significant ($F(38, 662) = 1.59$, $p=0.02$). However, the effects are less significant (adjusted $R^2=0.01$). Further examination of the time-fixed effects shows that the time-related effects are positive and exhibit an increasing trend (Fig. \ref{fig:time}.a). This indicates that the number of new commits is less associated with elite developer activities in the later phases of the project. However, for Model P2, where the bug cycle time in each project-month is the dependent variable, the time-fixed effects model is not significant ($F(38, 662) = 1.80$, $p<0.01$). The effects are very small (adjusted $R^2$ = 0.01). The time-related effects have slightly patterns (Fig. \ref{fig:time}.b).

For Model Q1, where the number of new bugs in each project-month is the dependent variable, the time-fixed effects model is significant ($F(38, 662) = 3.29$, $p<0.001$). The results are similar to the first one (Fig. \ref{fig:time}.a). The time-related effects are positive and increasing in general (Fig. \ref{fig:time}.c), indicating the impact of elite developers' activities on the number of the new bugs reported is shrinking over time. For Model P2, where the bug fix rate in each project-month is the dependent variable, the time-fixed effects model is not significant ($F(38, 662) = 1.07$, $p=0.36$). No meaningful effect could be detected (adjusted $R^2=0.00$). The time-related effects may be irrelevant to this quality indicator (Fig. \ref{fig:time}.d).

The above analyses reveal that: although the time-related effects are significant, the project-specific fixed effects models are much stronger than the time-related effects for all dependent variables.

\vspace{1mm}\noindent
\textbf{Answers to $\mathbf{RQ_3}$.}
Based on the above results, we can answer $\mathbf{RQ_3}$ as follows:

\begin{framed}
\noindent \emph{Elite developers' effort distributions have significant correlations with project outcomes. \begin{enumerate}
    \item Project Productivity: (a) Efforts on communicative and supportive activities are negatively correlated with the project productivity in terms of the number of new commits in each project-month (productivity loss); (b) Efforts on organizational activities are positively correlated with project productivity in terms of the the average bug cycle time in each project-month; however, efforts on organizational activities have much stronger negative effects. The overall effects are negative (productivity loss).
    \item Software Quality: (a) Efforts on organizational and supportive activities are positively correlated with the number of newly-found bugs in each project-month (quality loss); (b) Efforts on communicative activities are negatively correlated with the bug fix rate in each project-month; however, efforts on supportive activities have positive effects. Combine them together, the overall effects are likely to be positive (quality gain).
    \item Except for the bug fix rate, time effect analyses show that the impacts exhibit some decreasing trends with the progress of the project, which may result from the increasing proportion of non-elite developers' contributions in the latter stages of the project.
    \item In general, compared with the company-sponsored projects, effort distributions' correlations with project outcomes are less significant for non-company-sponsored projects.
\end{enumerate}}
\end{framed}

\section{Discussion}
\subsection{Discussions of the Findings}
\label{sec:discussionoffindings}

First of all, our results and findings confirm the important roles of elite developers in open-source development. As the results of $\mathbf{RQ_1}$ shows, they engaged in the majority of the projects' activities, though they only account for a small proportion of contributors in the entire community. Except for communicative activities, elite developers account for over 50\% activities in all the other three categories. The results confirm prior literature dating back to early 2000s \cite{Mockus:2002:TCS:567793.567795, Crowston:2008:FOS:2089125.2089127}. We can conclude that open-source projects are still largely driven by a small number of elite members after over 20 years of evolution. While such high concentrations may ensure bottom-line project outcomes, such situations may not be optimal for long-term health of a project \cite{crowston2006assessing}.
Engaging the non-elite users' participation through mechanism and technology innovation is still a challenge \cite{steinmacher2015social}.

Secondly, the results and findings of $\mathbf{RQ_2}$ show that the shifting of elite developers' activities did happen in most of the sampled projects. The activity shifting indicates the elite developers' role transitions with the growth of the project and the community. Organizational behavior theorists often argue that such transitions may be risky and troublesome for both individuals and organizations \cite{ashforth2000role, nicholson1984theory}. Let us imagine a situation that an elite developer may be involved. She used to enjoy the work of making technical contributions by committing high-quality code, but gradually, she finds herself having to spend more and more time on supportive work and communicating with novice users. This may conflict with her career goal. Unfortunately, at least in the software engineering community, this has not received any attention. Future research is necessary to address the issues related to such role transitions.

\sloppy
The results and findings of $\mathbf{RQ_3}$ reveal relationships between elite developers' effort distributions and project outcomes. In general, there are some negative associations. For three out of four project outcome indicators ($NewC_{im}$, $BCT_{im}$, and $NewB_{im}$) our results suggest putting more efforts into communicative, organizational, and supportive work is negatively correlated with the project outcomes. Elite developers are humans who have limited time and attention resources every day. If the three types of non-typical activities occupy too much of their time and attention resources, they may not be able to guarantee the productivity and quality of their contributions to technical tasks. Meanwhile, to fill such a gap, non-elite developers may have to contribute more in the development tasks. Since those non-elite developers often do not have a comparable level of technical expertise, their code could be more buggy, thus may lead to lower software quality \cite{aberdour2007achieving}. However, for the last project outcome indicator ($BFR_{im}$), our results show that the elite's efforts in supportive work do have positive correlations with project quality. A possible explanation is the efforts in supportive activities does help to maintain a good defect removal process, thus improve the bug fix rate in each project-month.

$\mathbf{RQ_3}$'s findings, if put together, describe a dilemma that elite developers often have to face in their projects. With the growth of their projects, they need to spend more time on non-technical tasks, which force them to reduce their technical contributions. Since their technical activities still account for a majority of the project's typical development work (see Tab. \ref{statisticsAC}), the project would also experience some productivity and quality loss. But doing more non-technical work is not meaningless: it perhaps helps to maintain a project's work processes (e.g., defect removal process) and is paid back by some quality gains.

Another finding worth noting is the differences between non-company-sponsored projects and company-sponsored projects. $\mathbf{RQ_3}$'s results indicate that company-sponsored projects tend to be more influenced by their elite developers' effort distributions. This is not surprising; such projects often rely on a small amount of full-time employees as the elite developers. Some of them may lack the interests to make voluntary contributions \cite{lerner2002some} and work a regular 8-hour daily schedule from 9 to 5. In case that non-technical work occupies more time, they do not use their own time to make up for the technical work. 

To sum up, our work does not only confirm the empirical observations of developers' activities in open source communities but also provides new findings and insights that shed light on future research. For example, we observe the elite developers' role transitions from the shifting of their work concentrations. Thus, supporting such transitions has not yet been investigated. Besides, we identify the impacts of effort distributions over the four broader categories on project outcomes. As far as our best current knowledge, it is the first piece of empirical evidence on this topic. How to leverage the findings to bring better project outcomes also requires follow-up research.

\subsection{Practical Implications}

Our findings suggest immediate practical implications. First, for most of the projects in our sample, the increase of elite developers often fails to keep pace with the growth of projects. This leads to heavy burdens to the elite developers. Indeed, many open-source projects seem to be too conservative to guarantee a member the permissions to perform some administrative tasks. While the open-source ideology is pretty progressive, its management structures are perhaps somewhat pre-industrial, i.e., a very small amount of elites share most of the authorities and powers in the community \cite{collier2010promoting, shah2006motivation, von2006promise}. Decentralizing such authorities and powers, particularly that related to routine work, might be a choice. It does not only alleviate elite developers' burdens but also give ordinary members in communities some extra motivations \cite{roberts2006understanding}. Besides, because the turnover of the core reviewers is high and rapid \cite{van2017reviewing}, allowing some non-elite developers to share some elite developers' routine duties would help to offset the negative impacts of their turnovers. 

Second, the differences between company-sponsored and non-company-sponsored projects indicate that the company-sponsored projects more or less inherit the management practices of the corporate world. Elite developers' involvements in non-technical tasks influence project outcomes in a more significant way. It seems that the elite developers tend to be trapped more on routine works. In his dissertation \cite{wagstrom2009vertical}, Wagstrom has shown that the vertical integration between companies and open-source communities would inevitably lead to increases in unnecessary communicative and organizational practices. Given the limited time and attention resources of developers, these unnecessary non-technical practices may hurt a project's productivity. Thus he recommended focusing on communication ``meeting individual coordination requirements.''  According to our results, his recommendation is still valid. Besides, from a company's perspective, avoiding ``copying'' their internal governing structures may be necessary even for the projects they dominate \cite{germonprez2016theory,schaarschmidt2015firms}.

\subsection{Design Implications}

With the growth of the project, elite developers often have to give more effort to communicative and supportive tasks. Our study reveals such a shifting of work may have negative impacts on project outcomes. As we discussed before in Section~\ref{sec:discussionoffindings}, these tasks are often necessary and cannot be ignored, building software tools to assist or partially free elite developers may be a good solution. 

Building such tools are feasible. At least for many organizational and supportive activities, there are technologies readily available. For instance, \emph{Assigned} and \emph{Unassigned} are two main events in the organizational activity category (see Fig. \ref{taxonomy}). The main time cost for them is to identify the assignee. These tasks can be easily automated with tools \cite{Anvik:2006:FTB:1134285.1134336}. The supportive work can be divided into two sets---maintenance and documentation. Let us have a look at maintenance activities first. For many raw activities associated with maintenance, there are ready-to-use automated tools built by researchers. For example, the \emph{CreateTag} can be automated using techniques such as \cite{Chaparro:2017:DMI:3106237.3106285}. Automatic subscribed and unsubscribed can be realized through learning users' characteristics \cite{bissyande2013got}. For documentation tasks, there are many metric-based or machine learning techniques ready for use \cite{luciv2018detecting, zesch2012text}, thus automating some \emph{MarkedAsDuplicated} and \emph{UnMarkedAsDuplicated} tasks.

Current technologies may be less mature for helping elite developers on communicative tasks. As shown in Fig. \ref{taxonomy}, communicative category contains four raw \textsc{GitHub} activities: \emph{Mentioned}, \emph{CommentDeleted}, \emph{IssueComment}, and \emph{CommitComment}. For some specific activities related to \emph{Mentioned}, researchers have developed techniques for automating them. For example, when mentioning somebody to fix an issue, the bug-fixer recommendation technique developed by Kim et al. \cite{kim2013should} may be directly applied to identify the target of the mentioning. Besides, \emph{CommentDeleted} tasks can be automated. For example, a disruptive message by a member can be automated deleted by a \textsc{GitHub} bot app equipped with advanced sentiment analysis techniques. Building automated tools for \emph{IssueComment} and \emph{CommitComment} requires some advanced techniques on abstractive semantic summarization and text generation, which are far from mature even in Natural Language Processing community \cite{thomsontoward2015, wen2015semantically,li2016diversity}. 

While there are many available techniques, most (if not all) of them have never been used by practitioners. This may be because such techniques have not been integrated into elite developers' normal workflow. As Terry Winograd and his colleagues \cite{winograd1986understanding} pointed out in their influential book ``\emph{Understanding computers and cognition: A new foundation for design}'', a computing application must be integrated to users' workflow in a non-intrusive way to gain widespread use.  

\subsection{Recommendations}

Our findings and the above discussions can be summarized into recommendations for practitioners and researchers.

Recommendations for open source practitioners are:

\begin{itemize}
    \item \emph{Open source projects may consider decentralizing the administrative authorities and powers related to routine tasks.} 
    \item \emph{Project members should focus on communication ``meeting individual coordination requirements.''}
    \item \emph{Projects sponsored by companies should avoid copying their sponsors' internal governing structures.}
\end{itemize}

We can also consider future research (incl. tool design and implementation) efforts with the following possible challenges.

\begin{itemize}
    \item \emph{Further understanding of developer activities.}
    \item \emph{Mechanism design for broadening participation in and sharing non-technical responsibilities.}
    \item \emph{Tool support for relieving elite developers from routine administrative burdens by synthesizing existing techniques to their routine workflow.}
\end{itemize}

\subsection{Threats to Validity}
As any empirical studies, our study is not free of threats to validity. We briefly discuss them from three perspectives.

First, from the perspective of \textbf{construct validity}, we are confident that there is no significant threat. Our study involves six primary constructs, which are four categories of \textsc{GitHub} activities, and project productivity and quality (each with two metrics, total four metrics). All their definitions and operationalizations are based on prior literature. For the four activity categories, we follow the standard procedure to develop the mappings between raw \textsc{GitHub} activities and these categories. The two project outcomes are adapted from literature; and each of them is measured by two distinct indicators. By using multiple indicators for one project outcome construct, our study does not only avoid to oversimplify the concept of ``productivity'' and ``quality'', but also brings more insights. Thus, we have the confidence that most of the threats to construct validity have been removed. 

Second, from the perspective of \textbf{internal validity}, we took multiple measures to ensure that the data collection process avoids the most of perils summarized in \cite{bird2009promises,kalliamvakou2014promises}. For example, all subjected projects are all large ones with established governing structure and practices, and use pull requests to manage members' contributions. The data used in the study are objective human activity records collected from online repositories. The analysis processes are unbiased. We use mature, widely-used analysis techniques, and empirically justify the use of the fixed effects models in panel regressions. 

One potential threat is that using \textsc{GitHub} data only. But doing so has its methodological justifications. While we acknowledge that the development trace data could be in multiple other channels such as email, IRC, forums, and so on, an unfortunate fact is that not all of them are publicly available. In fact, for the 20 projects studied in this paper, none of them has all the channels data ready. If we use multiple data sources for some projects but a single data source for the rest, guaranteeing the fair comparisons among them could be impossible. Moreover, using multiple data sources selectively would pose serious threats to the ``construct validity'' because establishing the mapping between activities and categories would require different protocols when crossing data sources. Thus, but weighed the gain and loss of using multiple data sources, we decide to use \textsc{GitHub} data only. At least, it guarantees the consistency at the methodological level, which is a basic requirement for any scientific inquiry \cite{bordens2002research,denscombe2014good,kothari2004research}. Thus, we view that not using multiple data sources as a limitation but not a serious threat to internal validity, as pointed out by Margaret-Anne Storey in her ICSE'19 keynote \cite{Storey:2019:PPQ:3339663.3339666}.

Third, from the perspective of \textbf{external validity}, we admit that our results may not be able to be generalized to all open source projects. However, the sampled projects represent a wide range of projects regarding the application domains. They also form a balanced sample of non-company-sponsored and company-sponsored projects. One potential limitation is that all 20 projects are large ones. We urge caution, however, for applying our findings in the context of small or medium size open source projects.

\section{Conclusion}
While elite developers' important role in open source development has been long known in software engineering literature, their activities have not been yet thoroughly investigated. Using fine-grained event data of 20 open source projects, our study paints a dynamic panorama of elite developers' activity, as well as their activities' impact on project outcomes in terms of project productivity and product quality.  

Our study yields a set of findings. First, our study confirms the essential roles of elite developers. Their activities account for the majority across all four types of broader activity categories: communicative, organizational, supportive, and typical. Second, our study reveals that elite developers' activities shift to the ``project management'' tasks from ``technical'' work. We observe that communicative and supportive activities increase much faster than typical development activities. Third, elite developers' effort distributions have significant correlations with project outcomes on productivity and quality. When they put more efforts into communicative and supportive work, a project's productivity (measured by the number of new commits ($NewC_{im}$) and bug cycle time ($BCT_{im}$) in each project-month) is likely to decrease. Besides, a project's quality (measured by the number of new bugs in each project-month ($NewB_{im}$)) is negatively associated with their activities on organizational and supportive tasks. But its another indicator---bug fix rate in each project-month ($BFR_{im}$)---is positively correlated with efforts in supportive activities, thus may increase when the elite developers put more efforts into such type of activities. These findings reveal a complicated picture of elite developers' effort distributions, and also partially indicate a dilemma faced by many OSS elite developers, i.e., with the growth of a project, its elite developers have to conduct more communicative and supportive work. We discuss the practical and design implication of the study.  

For future work, we plan to continue the focus on elite developers. We plan to replicate this study with a larger sample of projects and go one step further to explore the contextualized, individual differences among elite developers. Currently, the analysis unit is at the project-level, we also plan to extend the study by performing analyses at multiple levels, e.g., at the individual-level or the ecosystem-level. Moreover, project outcomes are much broader than productivity and quality. We plan to explore some alternative project outcomes, particularly those related to social and human development (e.g., the growth of newcomers). We will also design and implement tools to free (at least partially) elite developers from increasing communicative and supportive tasks, allowing them to maximize the impacts of their technical leadership in projects.

\begin{acks}
This work is partially supported by National Science Foundation under awards CCF-1350837 and IIS-1850067. 
\end{acks}

\bibliographystyle{ACM-Reference-Format}
\bibliography{sample-base}


\end{document}